\documentclass[twocolumn]{aastex63}

\usepackage{enumerate} 

\newcommand{\Vsig}{$(V/\sigma)_{R_{\rm{e}}}$}
\newcommand{\Ve}{$V_{R_{\rm{e}}}$}
\newcommand{\se}{$\sigma_{R_{\rm{e}}}$}
\newcommand{\re}{$R_{\rm{e}}$}
\newcommand{\spel}[1]{\textcolor{black}{#1}}

\usepackage{amsmath}
\shorttitle{LEGA-C Jeans Models}
\shortauthors{van Houdt et al.}

\begin{document}

\title{Stellar Dynamical Models for 797 $z\sim 0.8$ Galaxies from LEGA-C}

\correspondingauthor{Arjen van der Wel}
\email{arjen.vanderwel@ugent.be}

\author{Josha van Houdt}
\affil{Max-Planck Institut f\"{u}r Astronomie K\"{o}nigstuhl, D-69117, Heidelberg, Germany}

\author{Arjen van der Wel}
\affil{Sterrenkundig Observatorium, Universiteit Gent, Krijgslaan 281 S9, 9000 Gent, Belgium}
\affil{Max-Planck Institut f\"{u}r Astronomie K\"{o}nigstuhl, D-69117, Heidelberg, Germany}

\author{Rachel Bezanson}
\affil{University of Pittsburgh, Department of Physics and Astronomy, 100 Allen Hall, 3941 O'Hara St, Pittsburgh PA 15260, USA}

\author{Marijn Franx}
\affil{Leiden Observatory, Leiden University, P.O.Box 9513, NL-2300 AA Leiden, The Netherlands}

\author{Francesco D'Eugenio}
\affil{Sterrenkundig Observatorium, Universiteit Gent, Krijgslaan 281 S9, 9000 Gent, Belgium}

\author{Ivana Barisic}
\affil{Max-Planck Institut f\"{u}r Astronomie K\"{o}nigstuhl, D-69117, Heidelberg, Germany}

\author{Eric F.~Bell}
\affil{Department of Astronomy, University of Michigan, 1085 South University Ave., Ann Arbor, MI 48109, USA}

\author{Anna Gallazzi}
\affil{INAF-Osservatorio Astrofisico di Arcetri, Largo Enrico Fermi 5, I-50125 Firenze, Italy}

\author{Anna de Graaff}
\affil{Leiden Observatory, Leiden University, P.O.Box 9513, NL-2300 AA Leiden, The Netherlands}

\author{Michael V.~Maseda}
\affil{Leiden Observatory, Leiden University, P.O.Box 9513, NL-2300 AA Leiden, The Netherlands}

\author{Camilla Pacifici}
\affil{Space Telescope Science Institute, 3700 San Martin Drive, Baltimore, MD 21218, USA}

\author{Jesse van de Sande}
\affil{Sydney Institute for Astronomy, School of Physics, A28, The University of Sydney, NSW, 2006, Australia}
\affil{ARC Centre of Excellence for All Sky Astrophysics in 3 Dimensions (ASTRO 3D)}

\author{David Sobral}
\affil{Department of Physics, Lancaster University, Lancaster LA1 4YB, UK}

\author{Caroline Straatman}
\affil{Sterrenkundig Observatorium, Universiteit Gent, Krijgslaan 281 S9, 9000 Gent, Belgium}

\author{Po-Feng Wu}
\affil{National Astronomical Observatory of Japan, Osawa 2-21-1, Mitaka, Tokyo 181-8588, Japan}

\begin{abstract}
We present spatially resolved stellar kinematics for 797 $z=0.6-1$ galaxies selected from the LEGA-C survey and construct axisymmetric Jeans models to quantify their dynamical mass and degree of rotational support. The survey is $K_s$-band selected, irrespective of color or morphological type, and allows for a first assessment of the stellar dynamical structure of the general $L^*$ galaxy population at large lookback time.  Using light profiles from Hubble Space Telescope imaging as a tracer, our approach corrects for observational effects (seeing convolution and slit geometry), and uses well-informed priors on inclination, anisotropy and a non-luminous mass component. Tabulated data include total mass estimates in a series of spherical apertures (1, 5, and 10 kpc; 1$\times$ and 2$\times$\re), as well as rotational velocities, velocity dispersions and anisotropy.  We show that almost all star-forming galaxies and $\sim$50\% of quiescent galaxies are rotation-dominated, with deprojected $V/\sigma\sim1-2$. Revealing the complexity in galaxy evolution, we find that the most massive star-forming galaxies are among the most rotation-dominated, and the most massive quiescent galaxies among the least rotation-dominated galaxies.  These measurements set a new benchmark for studying galaxy evolution, using stellar dynamical structure for galaxies at large lookback time. Together with the additional information on stellar population properties from the LEGA-C spectra, the dynamical mass and $V/\sigma$ measurements presented here create new avenues for studying galaxy evolution at large lookback time.
\end{abstract}


\section{Introduction} 
\label{sec:intro}
The formation and evolutionary history of galaxies is encoded in the motions of their stars. The dynamical state of the Milky Way is now being revealed in ever more detail with Gaia (\citealt{GaiaDR2,GaiaDR2_kinematics}) and stellar kinematics of nearby galaxies has been a very active area of research since the 1970s (e.g., \citealt{Bertola1975,Binney1978,Kormendy1982,Davies1983}). In the present-day Universe, much has been learned from resolved kinematics measurements made with single slit observations (e.g. \citealt{TullyFisher,FaberJackson,FundamentalPlane1,Marel1991}). More recently, this has been superseded by integral field observations, revealing kinematic patterns in two dimensions (e.g. \citealt{Emsellem2004,Krajnovic2011}) and qualifying the dynamics of various types of galaxies (\citealt{Emsellem2011,Graham2018}). In particular, early-type galaxies are found to exhibit very diverse rotational patterns and are often split into fast rotators and slow rotators. Fast rotators are consistent with being axisymmetric, oblate rotators, whereas slow rotators are more dispersion dominated and often have complex dynamical structures (\citealt{Emsellem2007}). These observations have in turn sparked significant theoretical effort (e.g. \citealt{Jesseit2009,Bois2010,Bois2011,Naab2014,Lagos2018b,Lagos2018a,VdSande2019}), to understand the large variety in observed kinematic signatures and explain the varying degrees of ordered motion by investigating the relative importance of e.g. (minor) merging and accretion. \\
So far, a clear observational picture has not been produced at higher redshift. Yet it is highly desirable to understand the structure of high-redshift galaxies, when they were in the middle of forming the bulk of their stars and shortly thereafter. Ever following in the footsteps of studies of the local Universe, at higher redshift we have largely relied on photometry to guide progress. We have learned that the population of galaxies at $z\sim1$ is bimodal in color and can be separated into quiescent and star-forming galaxies based on optical colors (e.g. \citealt{Labbe2005,Williams2009}). Like in the local Universe, these groups show distinct structural properties (e.g. \citealt{bell12}) and correlations (e.g. \citealt{Franx2008,VanderWel2014a}). In particular, projected ellipticities have been used to study intrinsic shapes of galaxy populations which, for lack of more direct measures, have been associated with a galaxy's state of rotational support (\citealt{VdWel2014b,Chang2013}). Oblate, axisymmetric galaxies are found to form a larger fraction of the total galaxy population with increasing redshift, which is interpreted as the increased rotational support at high redshift.\\
Kinematic studies of ionized gas have revealed that large, star-forming disks exist at least as early as $z\sim 2$ (\citealt{ForsterSchreiber2006,Genzel2008}), but that those are more pressure-supported than gas disks in the present-day Universe (e.g., \citealt{Kassin2012a,Wisnioski2015}) .  Moreover, an increasing fraction of lower mass galaxies have not yet formed regular, rotating structures (e.g., \citealt{Forster2009}). This evolutionary trend has been successfully reproduced by hydrodynamical simulations (\citealt{Ceverino2014,Martig2014-II,Pillepich2019}).\\
However, there are limitations to the use of ionized gas as a tracer of dynamical structure: while to first order the integrated velocity dispersions of ionized gas and stars agree fairly well \citep{Bezanson2018b} there is large scatter for several reasons. First, it reveals only the state of the youngest parts of galaxies. Second, the spatial distribution of the emission is different from the total mass distribution: it has a patchy structure, which makes it a problematic tracer (\citealt{Varidel2019}).  The stellar component does not suffer from these drawbacks: they are the dominant mass component in most massive galaxies (within the inner 5-10 kpc for most galaxies in our sample), and the stellar light itself is a reasonable proxy for the shape of the mass distribution.\\
But the necessary data to access the stellar dynamical  structure  –  high-S/N continuum  spectroscopy – is challenging to obtain and have so far been limited to small samples (\citealt{DokkumMarel2007a,WelMarel2008,Newman2015,Newman2018,Guerou2017,Toft2017}) that only include very massive, passive galaxies. Even when these measurements can be made, the interpretation is complicated by the impact of observing conditions, such as the slit geometry and beam smearing. To make progress, detailed dynamical models can help.  At the expense of making assumptions about the underlying galaxies (e.g., axi-symmetry and regular motions) one can account for the impact of the large slit width, inclination and seeing  and  estimate  the  intrinsic  dynamical  properties  of galaxies.\\

The LEGA-C survey (\citealt{LEGAC2016}) is the first survey to provide resolved stellar kinematics for a magnitude-limited sample of galaxies at $z\sim0.8$. \citet{Bezanson2018} presented a first, qualitative look at the stellar rotation curves and dispersion profiles for a sample of $104$ quiescent galaxies in the LEGA-C survey. It was found that quiescent galaxies at $z\sim 0.8$, like their local counterparts, show a decrease of rotational support with increasing mass and that, like in the local population, there is a subset of massive galaxies seemingly devoid of significant rotation altogether. But crucially, the LEGA-C galaxies appeared to show significantly more rotation than local galaxies. However, the interpretation of the observed kinematics is complicated by the observational setup: the slit has a considerable width, the seeing is in many cases comparable to the galaxy size, and the slits are not necessarily oriented along the galaxy's major axis. An accurate quantitive description of the dynamical structure of these galaxies therefore requires dynamical modelling. \\
In this paper we present, for the first time, a concerted effort to infer the dynamical masses and dynamical states for a large, representative sample in the LEGA-C survey of quiescent and star forming galaxies. We achieve this by using axisymmetric Jeans models. The Jeans method is frequently used to model galaxies in the local Universe, for instance to constrain mass-to-light ratios and dark matter fractions (\citealt{Marel1991,Williams2009,Cappellari2013+jam}), or constrain the internal rotation and anisotropy of galaxies (\citealt{Satoh1980,Zhu2016}). Similarly, at $z>0$ they have been applied to study the dynamical masses (\citealt{Guerou2017}), initial mass functions (\citealt{Newman2017}) and rotation rates (\citealt{WelMarel2008,Newman2018}). \\
In this paper, we present the Jeans models for $797$ galaxies that are part of the third data release of the LEGA-C survey (van der Wel et al.~in prep.). In Section \ref{section:Section 2} we introduce the data and the sample selection used in this analysis. The models are described in Section \ref{section:Section 3}. In Section \ref{section:3.6} we show the results of the fits from the models to the observations, as well as a number of consistency checks. Some basic results from the derived quantities are shown in Section \ref{section:Section 5new}, after which Section \ref{section:Section 5} summarises and looks forward to anticipated follow-up work. Throughout the paper, a standard $\Lambda$CDM cosmology is assumed with $H_{0}=70$\,km\,s$^{-1}$, $\Omega_{\Lambda}=0.7$ and $\Omega_{b}=0.3$.

\section{Data}
\label{section:Section 2}
\subsection{LEGA-C Spectroscopy}
This work is based on the Large Early Galaxy Astrophysics Census (LEGA-C; \citealt{LEGAC2016}). LEGA-C is an ESO Public Spectroscopic Survey with the VIMOS instrument on the VLT/UT3. The survey targets galaxies at a redshift between 0.6 and 1.0 and below a redshift-dependent magnitude limit in the $K$ band, acting as a surrogate for a stellar mass selection yet insensitive to mass-modelling uncertainties. The survey has been completed and has collected 4081 spectra in the COSMOS field. This paper uses the full sample, to be published in the recent data release of the full sample \citep{van-der-wel21}.

The integration time is $\sim20$ hours per mask, resulting in spectra with a median signal-to-noise of $\sim19$ \AA$^{-1}$ at a spectral resolution of $R=3500$ (\citealt{straatman18}). Stellar kinematics are extracted with pPXF (\citealt{pPXF-2004,pPXF-2017}), as detailed in \citet{Bezanson2018}. In short, combinations of two sets of templates, for the stellar continuum and for the gas (line) emission, are fit to the observed spectra. The templates are broadened and shifted independently to optimise the fit in every row of the spectrum with $S/N>2$, giving us the line-of-sight velocity distribution (LOSVD) at each spatial pixel along the slit. Compared to \citet{Bezanson2018}, we use the pPXF fit with synthetic templates from C.~Conroy (priv. communication) instead of the empirical models based on the MILES (\citealt{MILES}) templates (\citealt{Vazdekis2010}). The Conroy templates offer increased resolution ($\sigma_{\rm{mod}} \sim 12~\rm{km~ s}^{-1}$ compared to $\sigma_{\rm{mod}} = 70~\rm{km~s}^{-1}$ for MILES) -- which is ideal compared to the resolution of our spectra ($\sigma_{\rm{instr}} \sim 35~\rm{km~s}^{-1}$). This high resolution is essential, as face-on disks in our sample can have stellar velocity dispersions as low as 40~$\rm{km~ s}^{-1}$.
We only measure the first and second moment of the LOSVD (the rotation velocity $v_{\star}$ and the observed velocity dispersion $\sigma_{\star}$, respectively) and do not fit for higher-order moments.

\subsection{Ancillary Imaging Data}
\label{Section:PhotometricCatalogs}
In addition to the resolved kinematics from LEGA-C, our dynamical models require a light tracer. We adopt for this purpose the HST/ACS F814W imaging from COSMOS (\citealt{COSMOS2007}), which covers nearly all galaxies in LEGA-C, and we use this imaging for fitting 2D light profiles with S\'ersic models as described by \citep{van-der-wel21}.

We also incorporate ground-based optical and near-infrared data from DR1 of the UltraVISTA survey (\citealt{UltraVista2013}), the photometric parent sample of LEGA-C. Using the $EAZY$ (\citealt{EAZY2008}) photometric redshift code in combination with the LEGA-C spectroscopic redshifts, we determine restframe $U-V$ and $V-J$ colours and select quiescent and star-forming galaxies according to the $UVJ$-criterion in \citet{Muzzin2013b}.

\subsection{Sample Selection for Jeans Modeling}
\begin{figure}
\epsscale{1.05}
\plotone{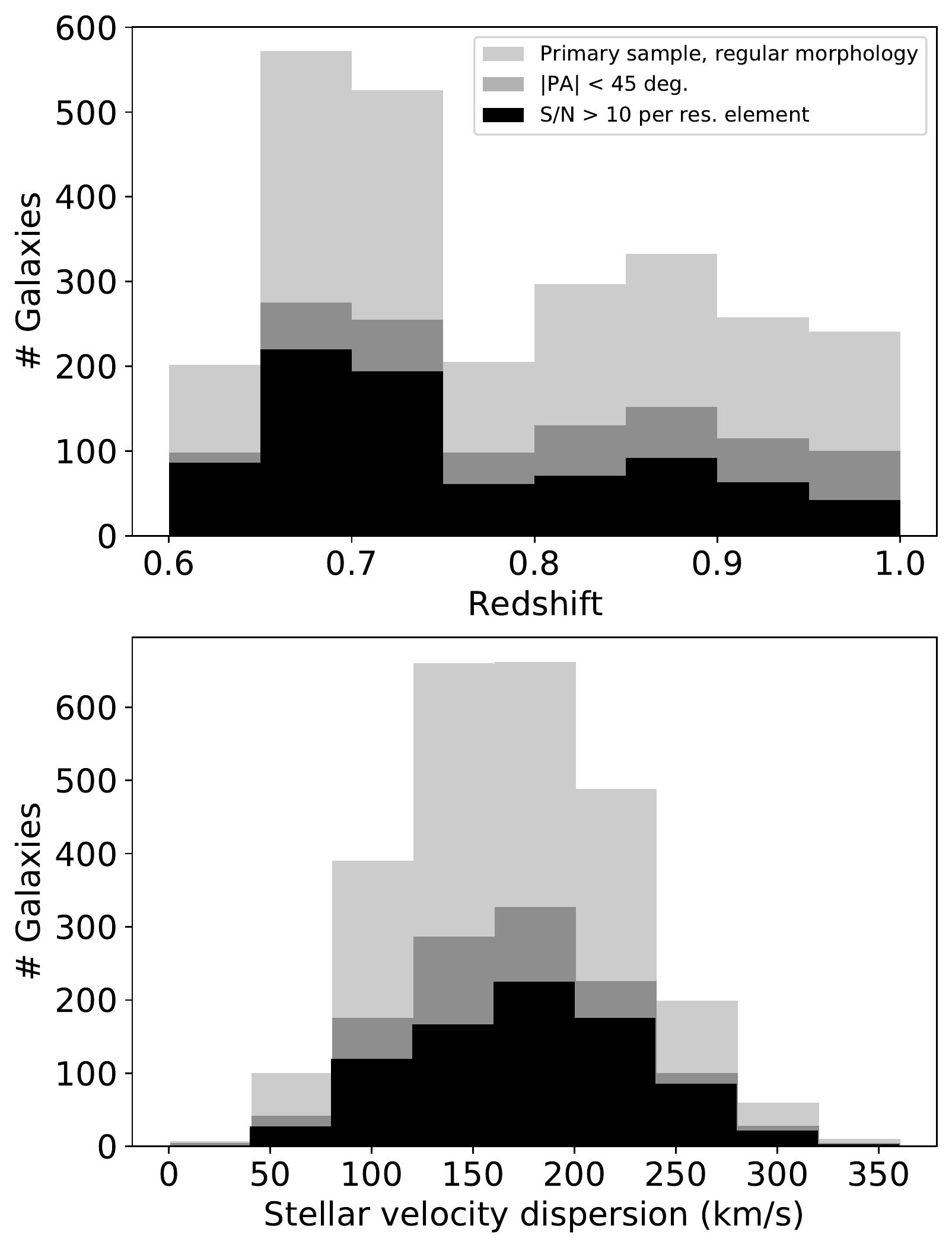}
\caption{Distribution of redshift (top panel) and integrated stellar velocity dispersion (bottom panel) for the parent sample of galaxies with regular morphologies and no flags (see text) in light grey, those with position angles aligned with the slit to within 45 degrees (darker grey) and the final sample of 861 spectra that also satisfy the $S/N$ criteria.\label{fig:sample}}
\end{figure}

\begin{figure}[h!]
\epsscale{1.1}
\plotone{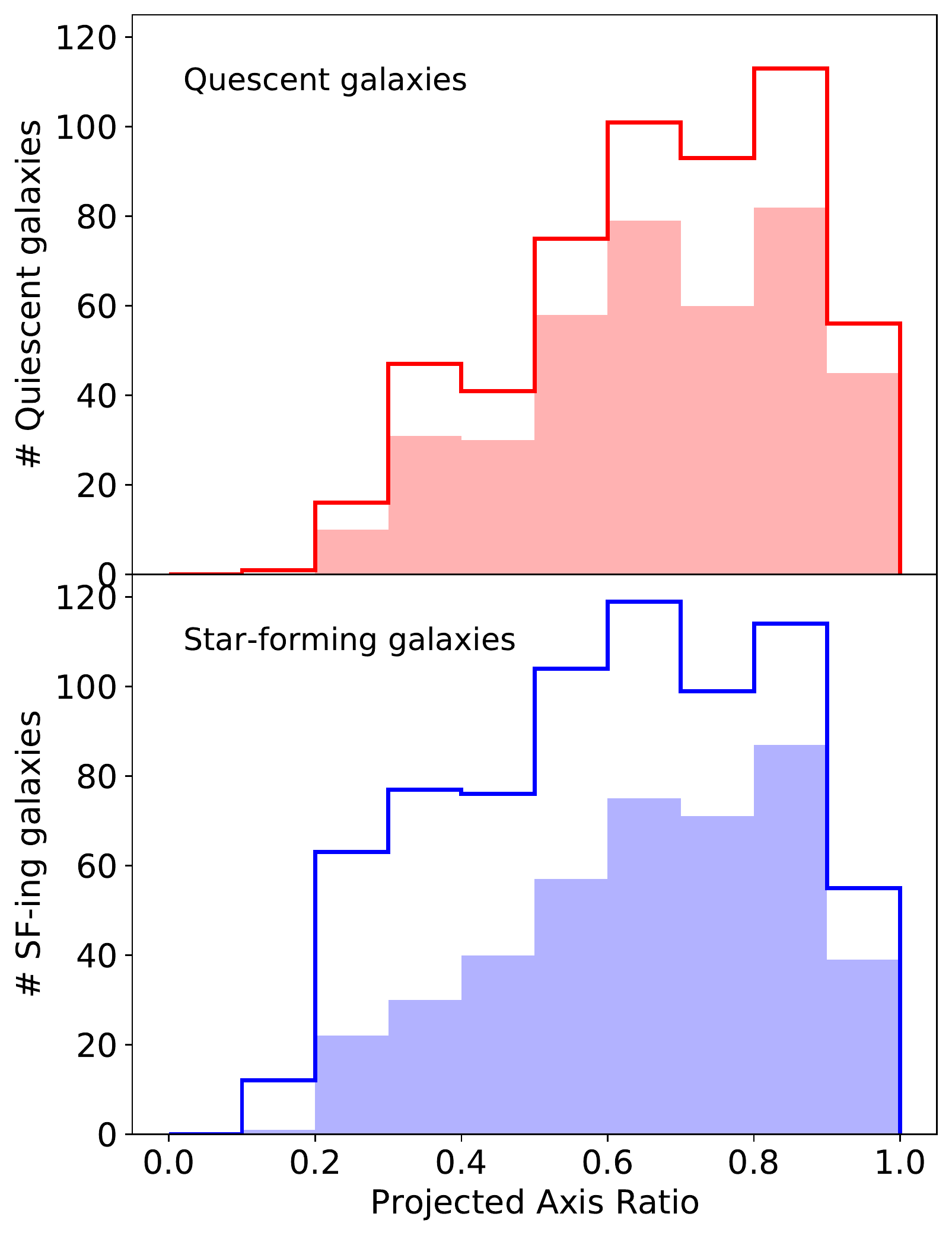}
\caption{Axis ratio distribution distribution of quiescent (top) and star-forming (bottom) galaxies. The lines indicate the eligible samples (primary targets with measured (integrated) stellar velocity dispersions and aligned with the slit to within 45 degrees); the solid histograms show the sub-samples with sufficient $S/N$ in the spatially resolved spectra for constructing the dynamical model. The $S/N$ for edge-on, star-forming galaxies suffers from attenuation, introducing an inclination-dependent bias in our sample of star-forming galaxies. This bias is absent for quiescent galaxies. \label{fig:qhist}}
\end{figure}

\begin{figure}
\epsscale{1.05}
\plotone{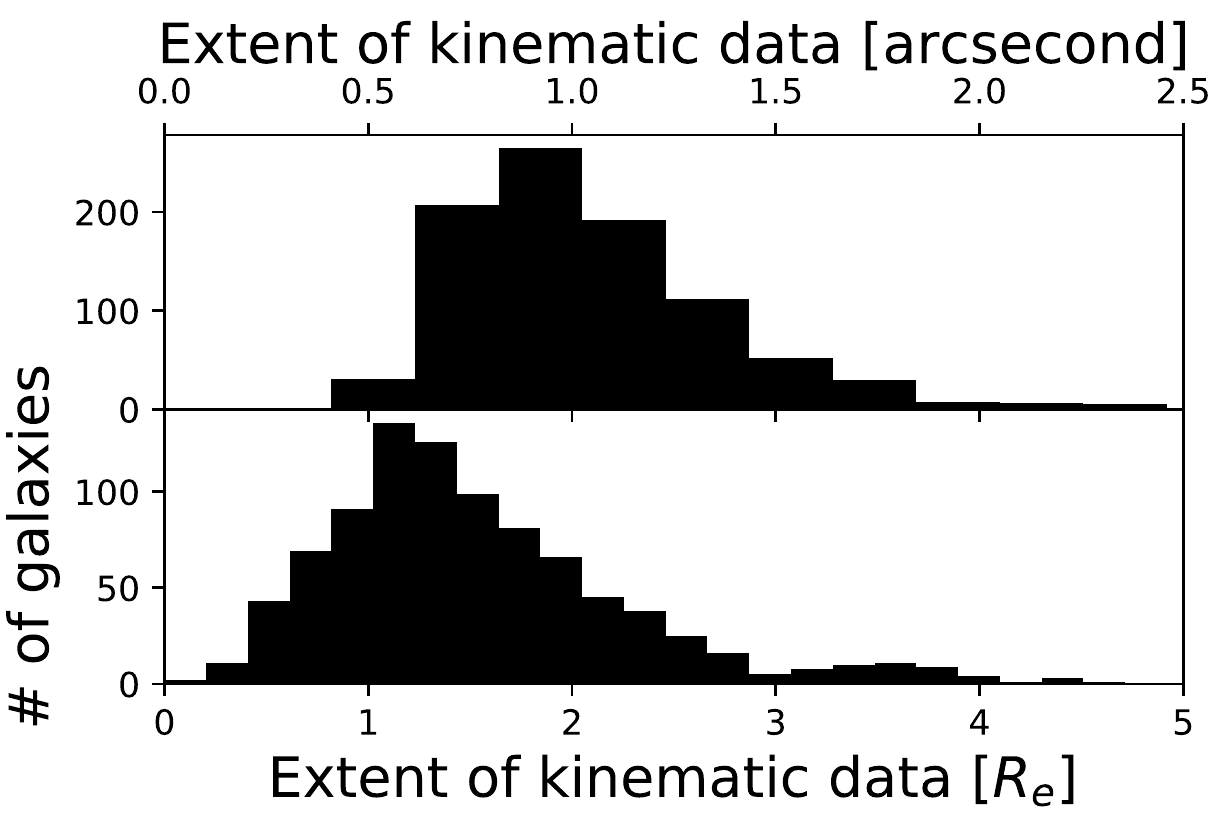}
\caption{Spatial extent of the kinematic measurements for $v_{rot}$ and $\sigma_{*}$ for the $860$ kinematic data sets (of 795 unique galaxies) in the final sample, first in arcseconds and then in multiples of each galaxy's effective radius.\label{fig:rmax}}
\end{figure}

The third (and final) LEGA-C release \citep{van-der-wel21} contains $3028$ spectra of $K_s$-band selected `primary' galaxies in the redshift range $0.6<z<1$. We  construct dynamical models for a subset of these, those 1) with the major axis is not misaligned with the slit by more than $45^{\circ}$ (1554 left), 2) without a morphology flag (mergers, irregulars, ...) or a spectroscopy flag (significant AGN contributions or severe flux calibration problems) as defined in the data release paper (1381 left), and 3) with $S/N>10$ in at least 3 spatial resolution elements (861 left).

Due to the observing strategy and the way in which galaxies are prioritized, some galaxies are observed in multiple masks. This means we have independent measurements of the same galaxy under slightly different seeing conditions and slit alignments. In the sample of 861 we have 797 unique galaxies: there are $64$ `duplicates' ($7.6\%$ of the sample), which we use for consistency checks of the method (Sec.~\ref{sec:duplicates}). 

Figure \ref{fig:sample} shows the distribution of $z_{spec}$ and $\sigma_{*}$ -- the integrated stellar velocity dispersion -- and how the selection criteria affect these distributions. 
Figure \ref{fig:qhist} shows the bias against edge-on star-forming galaxies: these galaxies have higher dust attenuation and often fall short of our $S/N$ criterion. This is an important consideration when attempting to quantify the relative numbers of flat and round star-forming galaxies, and the evolution thereof. The trend does not exist for the mostly dust-free quiescent galaxies. For those (and for face-on star-forming galaxies) the $S/N$ can be too low due to low stellar mass and/or high redshift. 

From the single-slit spectra we have one kinematic axis and a spatially resolved profile for $v_{\star}$ and $\sigma_{\star}$ for typically $\sim8-12$ spatial pixels, reaching over $2$ effective radii for some of the galaxies in our sample. Figure \ref{fig:rmax} shows the range probed for the galaxies in the sample, in arcsec as well as in units of the galaxy effective radius \re.\\

\section{Methods}
\label{section:Section 3}
\subsection{Axisymmetric Jeans models}
\label{label:JAM}
The dynamical models used in this work are based on the Jeans equations, a set of stellar hydrodynamic equations relating (average) kinematic properties to the stellar light and the underlying matter distribution. They are derived by taking velocity moments of the Collisionless Boltzmann Equation, describing the evolution of the phase-space density, assuming the system of particles to be collisionless and in steady-state (e.g. \citealt{BinneyTremaine}) \\
In this work we implement this using the Jeans Anisotropic MGE models (JAM; \citealt{JAM2008})\footnote{The JAM source code is available from \url{http://www-astro.physics.ox.ac.uk/~mxc/software/}}. This software uses the computationally convenient Multi Gaussian Expansion (described below) to model the surface brightness, the gravitational potential and the point spread function (PSF, see Sections \ref{section:masslight} and \ref{subsection:effective seeing}). The Jeans equations become solvable by making observationally-motivated assumptions on the velocity ellipsoid. Via the anisotropy parameter $\beta_{z}\equiv 1 - \langle v_{z}^{2}\rangle/\langle v_{R}^{2}\rangle$ (in cylindrical coordinates) we can model the flattening of the velocity ellipsoid in the direction perpendicular to the disk of the galaxy. The velocity ellipsoid is otherwise assumed to be aligned with the main cylindrical axes such that cross-terms ($\langle v_{z}v_{R}\rangle$, $\langle v_{\phi}v_{R}\rangle$, $\langle v_{\phi}v_{z}\rangle$) all vanish. This leaves us with one independent (possibly spatially varying) second velocity moment at every position. Given the inclination, this can then be integrated along the line-of-sight as done in Equation 28 of \citet{JAM2008}. This prediction can then be compared to the observed combination of 1st and 2nd order line-of-sight velocity moments $v_{rms} \equiv \sqrt{v^{2}_{\star} + \sigma_{\star}^{2}}$, allowing us to constrain the kinematic structure and gravitational potential with only a few parameters. \\
$v_{rms}$ allows us to constrain the \emph{total} rms motion in the galaxy as a sum of rotation and dispersion, but it gives no information about the relative importance of each of the components. To distinguish between these we make the additional assumption that the dispersion in the azimuthal direction is related to the dispersion in the radial direction such that
\begin{equation}\overline{v_{\phi}} = \kappa \sqrt{\overline{v_{\phi}^{2}}-\overline{v_{R}^{2}}}\end{equation}\label{eq:k}
%
where the constant rotation parameter $\kappa$ is a fitting parameter \citep{BinneyTremaine, DokkumMarel2007a}.

\subsection{Light and Mass Modelling}
\label{section:masslight}
An important input of the dynamical models is a detailed parametrisation of the tracer light (stellar), from which the observed kinematics originates, as well as the underlying gravitational potential. In our model, the potential is due to two mass components: the stellar body and dark matter. That latter, in actuality, represents any component that breaks the mass-follows-light assumption inherent in the former, but this will be discussed in full below.   

Both are described with a set of elliptical Gaussians using the Multi Gaussian Expansion code (\citealt{MGE2002})\footnote{The MGE software is also available at \url{http://www-astro.physics.ox.ac.uk/~mxc/software/}}. Extra Gaussian components are iteratively added until the fit does not improve significantly. All Gaussians are assumed to have the same, global PA (no isophotal twists). In addition, we impose a regularisation to limit the axis ratios around the global value as long as it does not affect the quality of the fit, which is most relevant if we add residuals or decompose mass profiles (see below). This is important since the minimum axis ratio limits the minimum allowed inclination in the dynamical fit. In some cases (in particular, for S\'{e}rsic profiles that fall off faster than a single Gaussian function, $n<0.5$) we also allow negative Gaussian components while ensuring that the sum stays positive.\\
\begin{figure}
\epsscale{1.3}
\plotone{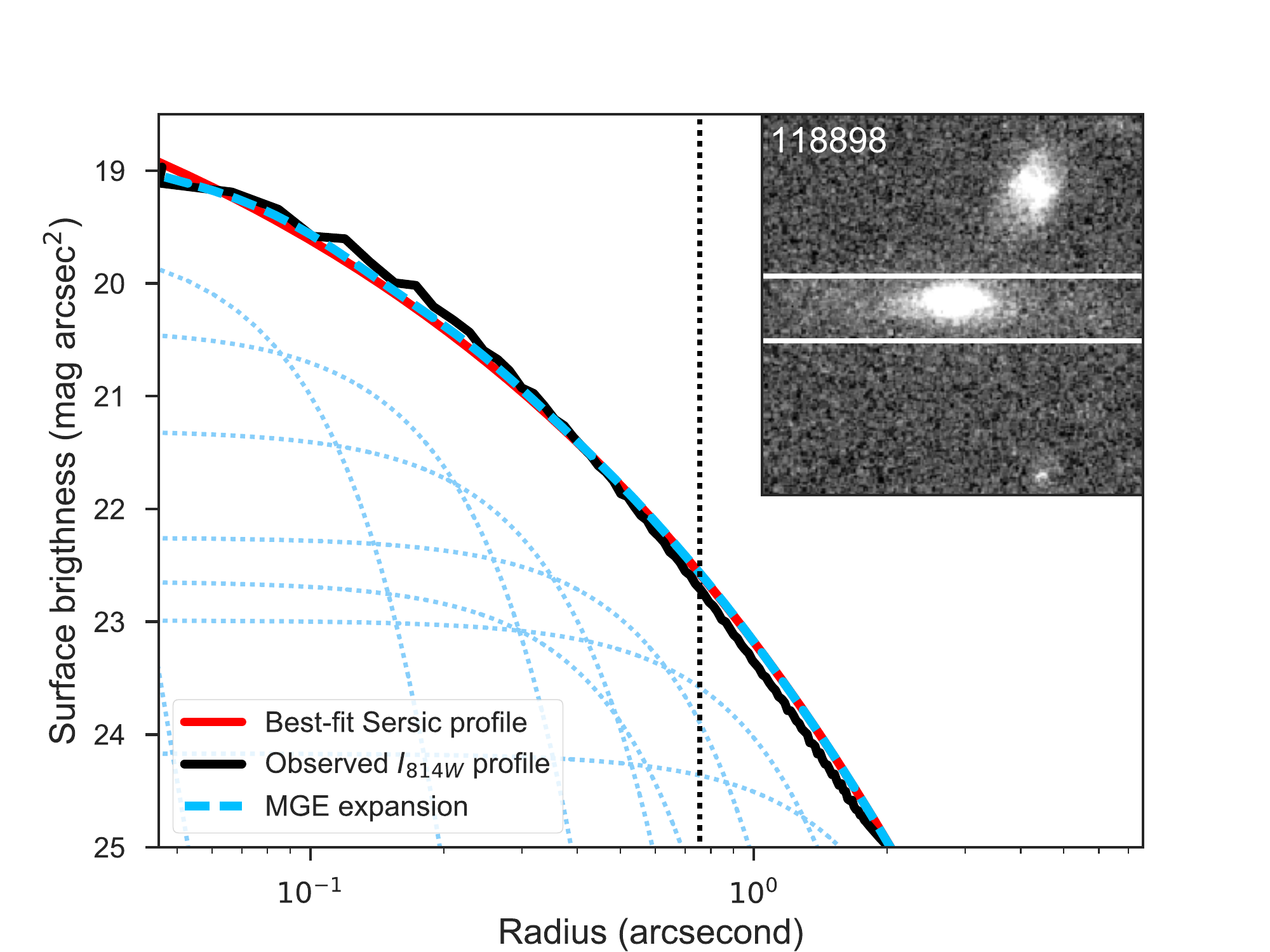}
\caption{Deconvolved surface brightness distribution -- used as light tracer input and stellar mass component in our dynamical model -- for the galaxy $1414$. The inset shows the $ACS/F814W$ image, with the VIMOS slit overlaid in white. The main panel shows observed surface brightness profile in black, as an average in elliptical annuli following the best-fit GALFIT geometry (PA and axis ratio). The solid red line is the best-fit S\'{e}rsic profile. The dashed blue line is the multi-Gaussian expansion of the S\'{e}rsic profile; individual Gaussian components are shown in dotted blue. Finally, the dotted vertical line denotes the effective radius of this galaxy.  \label{fig:mge_example}}
\end{figure}
\subsubsection{Light Tracer and Stellar Mass Model}\label{sec:lighttracer}
We use the ACS/F814W imaging as the light tracer model. The \emph{HST}/ACS instrumental point spread function has a non-negligible size relative to the galaxy sizes, which is why we use the best-fit S\'{e}rsic profiles to describe the deconvolved surface brightness distributions, following \citet{VdWel2012} and \citet{LEGAC2016}. These are then subsequently fitted with the MGE, as illustrated in Figure \ref{fig:mge_example}, where we show that this procedure accurately follows the true light profile over 6 magnitudes in surface brightness, out to 2.5 effective radii.

We note that for these relatively low-$S/N$ F814W data direct MGE fits systematically underestimate the fluxes, on average, 20\% as inferred from a comparison with ground-based photometry, where we synthesized F814W photometry from the SED fits to the ground-based photometry described in Section \ref{Section:PhotometricCatalogs}. Low surface brightness light is missed by the MGE, leading to underestimated effective radii. The S\'ersic model fluxes, on the other hand, show no such offset, justifying the extrapolation of the profiles out to large radii. The 1$\sigma$~scatter for both the MGE and S\'ersic fluxes is 15\% with respect to ground-based fluxes.

\subsubsection{Dark Matter Model}\label{sec:dark}

In addition to the luminous matter we also include a dark matter component. Later, in section 5.2, we will see that the primary role of the dark matter fit component is to allow us to fit mass to light variations within the inner 10kpc or so of the galaxy, having no predictive power outside of that radius. Consequently, while we choose a functional form for this component that is rooted in cosmology, its detailed functional form is not important and does not significantly influence the results of our overall fits. 

We adopt a spherically symmetric NFW-profile (\citealt*{NFW1997}):
\begin{equation}
\rho = \frac{\Delta \rho_{crit}}{(r/r_{s})(1+(r/r_{s}))^{2}}
\end{equation}
where $\rho_{crit}$ is the critical density, $r_{s}$ a characteristic scale and $\Delta$ the threshold level delimiting the halo, as a matter of convention often chosen to be $200$. The halo mass $M_{200}$ then corresponds to the mass inside the radius $r_{200}$ for which the average density is $200\rho_{crit}$. Furthermore, it is convenient to define the concentration parameter $c_{200}=r_{200}/r_{s}$. It is well known that there is a correlation between  halo mass and  concentration, such that more massive haloes tend to be less concentrated, though the scatter is large. From the simulations in \citet*{DuttonMaccio2014} we can use the relation between mass, concentration and redshift to eliminate one of the two halo parameters:
\begin{equation}
\log_{10}\left(c_{200}\right) = a+ b \log_{10}\left(M_{200}/[10^{12}h^{-1}M_{\odot}]\right)
\end{equation}
where the parameters $a$ and $b$ are given (for $\Delta=200$) by
\begin{equation}
a =  0.520+ \left(0.905-0.520\right)e^{-0.617z^{1.21}}
\end{equation}
and
\begin{equation}
b = -0.101 +0.026 z
\end{equation}
This allows us to reduce the dark matter component to a 1-parameter model: its mass $M_{200}$. These models can be easily included in the dynamical fit by expanding the NFW profile in a series of Gaussians and adding them to the MGE terms of the stellar matter.

We note that, in practice, this implementation means that the dark matter component always has the same slope (the inner slope from the NFW profile) and the only variable is the normalization, as dictated by $M_{200}$. The reason for this is that $r_s$ always lies beyond the spatial extent of our observations.  We stress again that we interpret the dark matter component \emph{not} as a constraint on the mass of entire dark matter halo, but as a constraint on deviations from the mass-follows-light assumption, which physically can be due to dark matter, but also contributions from gas and stellar $M/L$ variations.  Global differences in stellar $M/L$ are accounted for varying the $M/L$ of the light component, as are gas disks that follow the same profile as the stellar light. However, if gas or stellar mass profiles differ from the stellar light profile, then the `dark matter' component will attempt to account for this. The assumption of a spherical dark matter distribution is then not correct, but for the enclosed mass estimate this has a minor effect.

We note that we have to allow for a dark component with a negative mass.  This may be counter-intuitive, since setting a strictly positive prior seems physically self-evident.  The two reasons for allowing a negative mass here are 1) avoid a positive bias in the total mass, and 2) allow for inverted stellar $M/L$ gradients.  The positive bias occurs when the dark component is not constrained by the data, such that the median in the posterior is always positive. Allowing for negative masses improves the calculation of the uncertainties by symmetrizing the posterior around zero if there are no constraints on the dark component, avoiding the aforementioned bias. A side effect is that for a small set of galaxies we find a negative total mass (see Section \ref{sec:catalog}), which should be interpreted as unavoidable noise in the measurements.

\subsection{Effective Seeing}
\label{subsection:effective seeing}
We correct for beam smearing on an object-by-object basis. Thus, we wish to quantify the effective seeing of each galaxy, which reflects a combination of atmospheric seeing as well as observational and instrumental effects that broaden the light distribution. We do this by comparing the wavelength-collapsed LEGA-C spectra (equivalent to a radial light profile of that part of the galaxy covered by the slit) with a model light curve. The model light profile is obtained by convolving the ACS/F814W images successively with a seeing kernel and the slit. Then we rebin this to the $0.205\arcsec$ VIMOS pixel scale, and integrate the resulting image.

For the seeing kernel we use a Moffat profile (\citealt{Moffat1969}):
\begin{equation}M(\alpha,\gamma) = A\left(1+\frac{(x-x_{0})^{2}}{\gamma^{2}}\right)^{-\alpha}\end{equation}
where $A$ is the normalisation and $\gamma$ and $\alpha$ are the Moffat parameters we fit for. Given $\alpha$ and $\gamma$, it is straightforward to show that the full-width-half-maximum (FWHM) is given by $FWHM = 2\gamma\sqrt{2^{1/\alpha}-1}$. Generally, a Moffat profile has been shown to give a better description of the effect of atmospheric turbulence than a Gaussian function (\citealt{Trujillo2001}) because of the wings, controlled by the $\alpha$ parameter. We allow for a possible gradient in the background flux (due to e.g. imperfections in the spectroscopic data reduction) by adding a variable linear component to the Moffat-convolved model light profile. If $\alpha$ is poorly constrained we fix it at $\alpha = 4.765$ \citep{Trujillo2001} and only fit for $\gamma$. For objects where we did not find a good fit we use $\alpha=4.765$ and $\gamma$ derived from the mask-average FWHM and $\alpha$.\\
It is encouraging that despite atmospheric dispersion, drift, flexure and hysteresis the true PSF (typically, 0.75$\arcsec$) is in fact better than the DIMM seeing reported at the telescope (typically, 0.95$\arcsec$, \citealt{straatman18}). Interestingly, the variation from mask to mask is smaller than the variation from object to object within a given mask. This suggests that technical details of the instrument (focus, positional accuracy, alignment and location within the mask) are more important in determining the outcome of a spectrum than the variation of atmospheric conditions averaged over 20 hours of integration (good conditions were required to execute observations).

\subsection{Systematic Effects in Slit Spectroscopy}
\label{section:slit effects}
In general there are several things to keep in mind when using slit spectroscopy, in particular at the targeted redshift range where the slit width of 1$\arcsec$ is comparable to the effective radii of galaxies.

\subsubsection{Slit Misalignment}
During observations the slits are centered on the galaxies as accurately as possible, but this might still be prone to both random and systematic offsets. This will lead to an offset in the measured velocity profile, as well as asymmetries in the $v$ and $\sigma$ profiles if the slit is not aligned with the galaxy's major axis, even if the galaxy is perfectly axisymmetric. Although for a small subset of individual sources this offset can be determined (see Appendix A), finding constraints systematically for all sources was not possible. Nevertheless, the sources for which we \emph{can} constrain the offset suggest the slits are generally well-centered, with offsets of typically $0$, $1$, or sporadically more pixels, where 1 pixel corresponds to 0.205 arcsec. This motivates us to use a Gaussian prior on the slit center $x_{0}$ in the dynamical models, centered at $0$, with a standard deviation of $1.5$ pixels. 
\subsubsection{Slit Orientation}
\label{Section:ArtificialGradient}
\begin{figure}
\epsscale{0.7}
\plotone{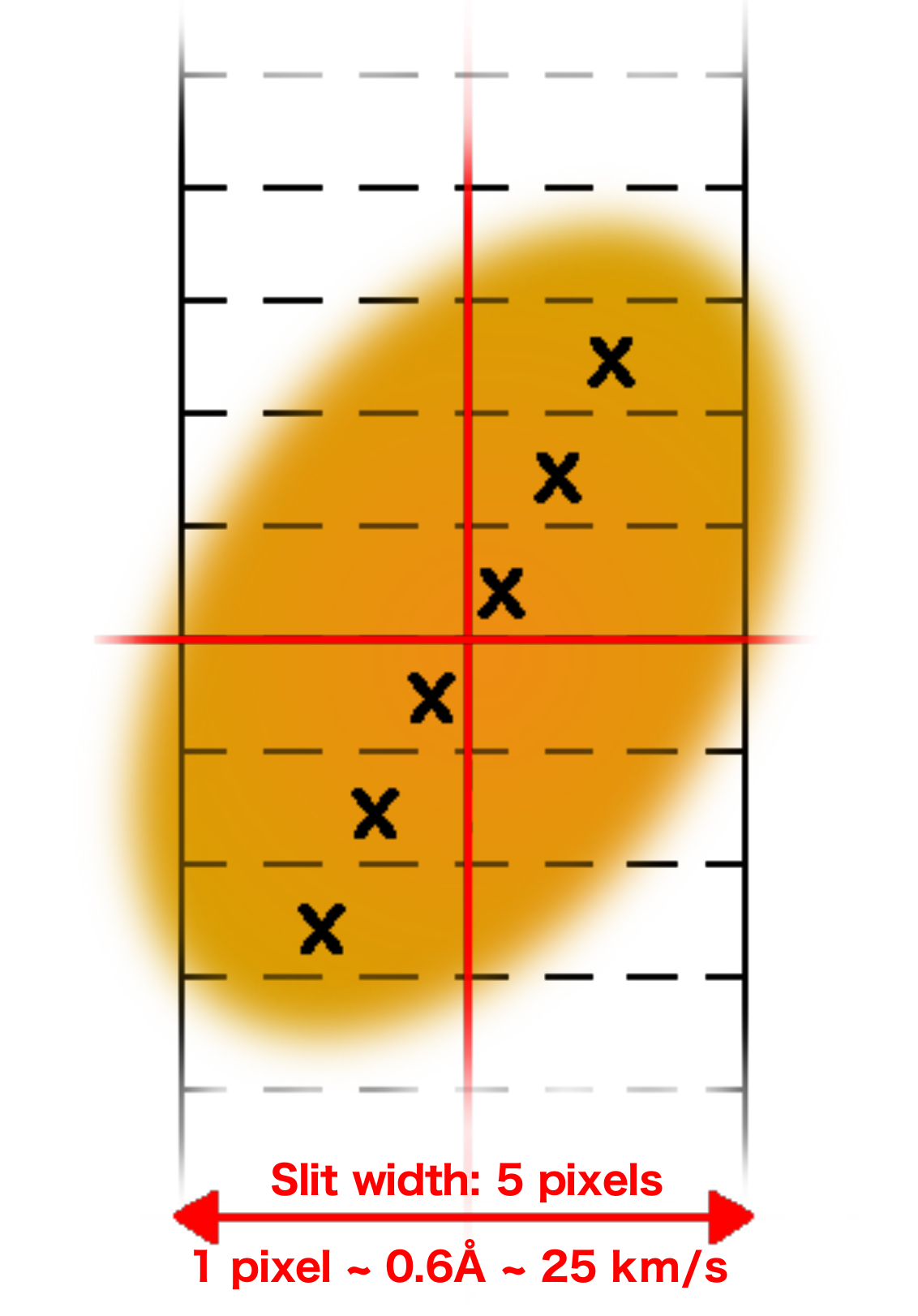}
\caption{Illustration of how an artificial velocity gradient arises in the slit. The black lines illustrate the slit, with dashed lines denoting the separation between spectral pixels. The orange blob illustrates the galaxy, the major axis of which is rotated with respect to the direction of the slit. Because of this, the center of the light of any individual pixel is not in the middle of the slit, but rather slightly shifted - indicated here with a black cross. The conversion of this spatial offset to an offset in wavelength and thus in an offset in derived velocity is indicated at the bottom (even though this varies slightly from galaxy to galaxy due to different spectral coverage.)\label{fig:slit_grad}}
\end{figure}
In deriving the rotational velocity at any particular spatial position, the light is tacitly assumed to come from a homogeneous light distribution that fills the entire 1-arcsec slit in the wavelength direction (this is how the spectral resolution $R$ in instrument manuals is usually defined). However, this is not true for galaxies that have light disributions that change substantially in wavelength direction of the slit, which is the case for stars or distant galaxies. This results in spurious velocity shifts if the major axis is misalignment with respect to the slit. We derive and subtract this offset as follows. First, we convolve the F814W MGE with the PSF. For every pixel, we then find the light-weighted center which we convert into a shift in wavelength, using the dispersion of $0.6$\AA \:per pixel. Using the median of the observed spectral coverage of each object, we then convert this into a shift in velocity. This is illustrated in Figure \ref{fig:slit_grad}. This artificial offset reaches up to $20$ km$\cdot$s$^{-1}$, being highest away from the center and for the flattest, most misaligned galaxies. \\

\subsubsection{Pixel Convolution}
To compare the JAM predictions to the observations, we need to convolve the model grid with a kernel that is a combination of the seeing PSF as well as the slit. The former is an MGE of the Moffat profile derived in Section \ref{subsection:effective seeing}. The latter is a top-hat function of 1 arcsec in width (the size of the slit in the wavelength direction).\\

\subsection{Inclination and Anisotropy Priors}
\label{section:3.5}

The observed kinematics depend on the inclination, but the only \emph{a priori} constraint we have is the observed axis ratio. If we know the distribution $f(p,q)$ of the \emph{intrinsic} axis ratios $p\equiv a/b$ and $q\equiv b/c$ of a given class of objects (e.g. quiescent galaxies in a given redshift range) we can construct the likelihood of observing the observed axial ratio $q^{\prime}=a^{\prime}/b^{\prime}$ for a given inclination. This will then serve as the prior on the inclination in the fits of the Jeans models to the total rms motion in the galaxy.\\
The intrinsic axis ratios $p$ and $q$ are related to the observed axis ratio $q^{\prime}$ through two viewing angles. In the oblate axisymmetric case (with $p=1$) this becomes a simple relation between the inclination, the observed axis ratio $q^{\prime}$ and the intrinsic short-to-intermediate ratio $q$:
\begin{equation}q^{\prime} = \cos^{2}{i} + q\sin^{2}{i}\end{equation}
We use the methodology described by \citet{Chang2013} to reconstruct the intrinsic $q$ distribution, assumed to be Gaussian, from an observed set of axis ratios, as shown in Figure 4.  For our sample we find a mean $\mu_q = 0.41$ and a scatter $\sigma_q = 0.18$.  This distribution allows us to derive a prior for the inclination: given an observed, projected axis ratio, what is the probability distribution for the intrinsic shape and, hence, the inclination?  The mathematical tools for this inversion are presented by \citet{van-de-ven21}.

While we see in Figure 4 that star-forming galaxies, on average, have flatter intrinsic shapes than quiescent galaxies \citep{Chang2013, VdWel2014b}, adopting different shape distributions as priors for the inclination would introduce a systematic difference in rotational properties for quiescent and star-forming galaxies, even if they were kinematically the same.  This spurious result would be explained, trivially, by the priors and not by the stellar kinematics.  In order to avoid this we adopt the same intrinsic shape distribution (that is, the same inclination prior) for different types of galaxies such that any difference in dynamical structure can be attributed to differences in the observed kinematics. As we will see, despite not assigning different priors to different types of galaxies, the model successfully distinguishes between rotating and non-rotating galaxies.

Given the assumed inclination, the anisotropy $\beta_{z}$ can in principle be well-constrained with 2D kinematics (\citealt{JAM2008}) but the long-slit data is expected to put weak constraints on the anisotropy. We therefore use a uniform prior of $-0.5<\beta_{z}<0.5$.

\begin{figure}
\epsscale{1.3}
\plotone{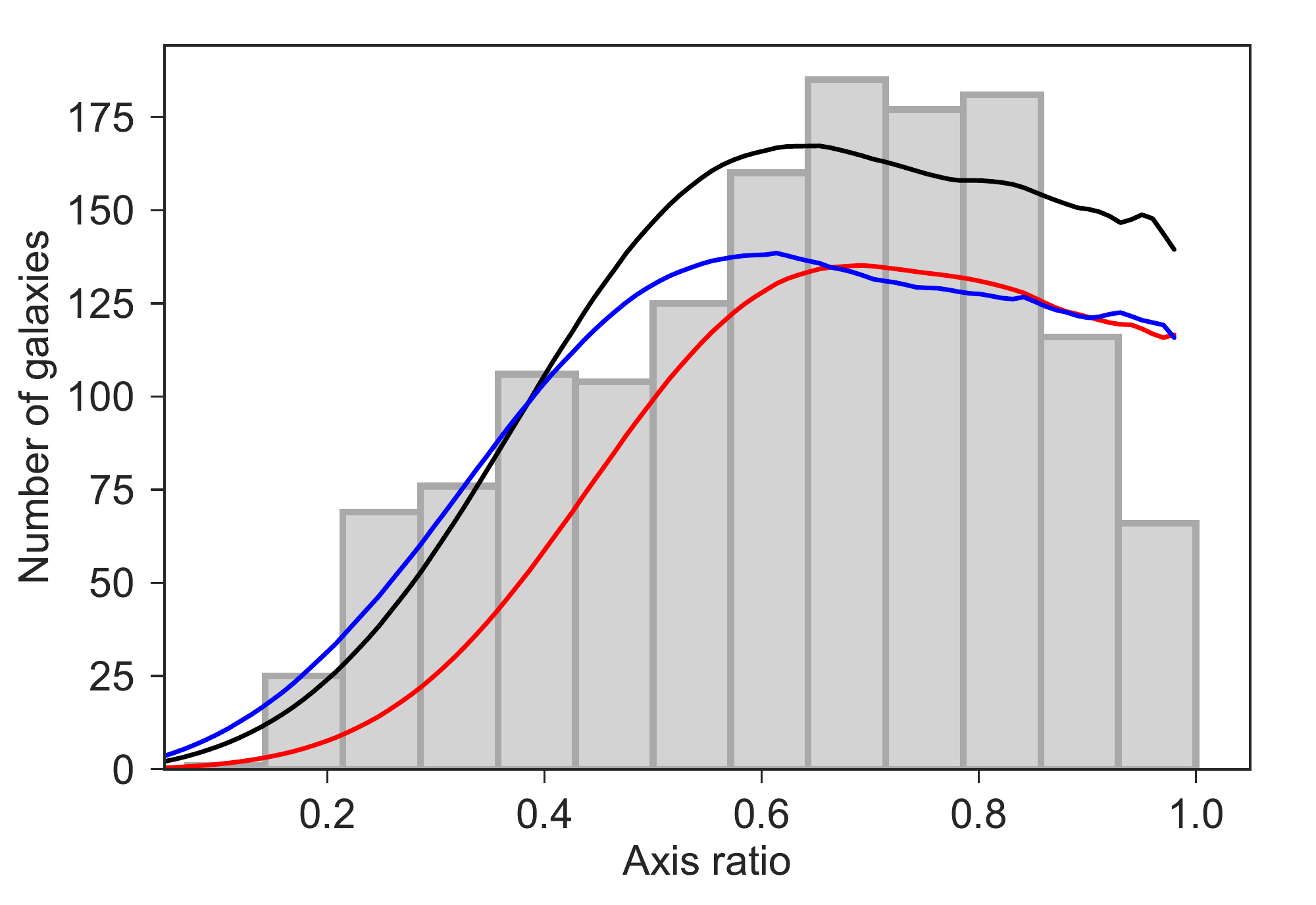}
\caption{The observed axis ratio distribution is fitted with a population of intrinsically oblate objects, viewed under random angles. The grey histogram is the observed axis ratio distribution for the whole primary LEGA-C sample; the black line is the best-fit oblate model. The red and blue lines are similar fits, but only to the star forming or quiescent galaxies, respectively. The joint fit is used in the remainder of the paper to describe the intrinsic shape distribution of the galaxy population, which, given the observed projected axis ratio of an individual galaxy, gives a prior probability distribution for the inclination in the dynamical model.\label{fig:axis_ratio_fit}}
\end{figure}

\begin{figure*}[p!]
\epsscale{1.1}
\plotone{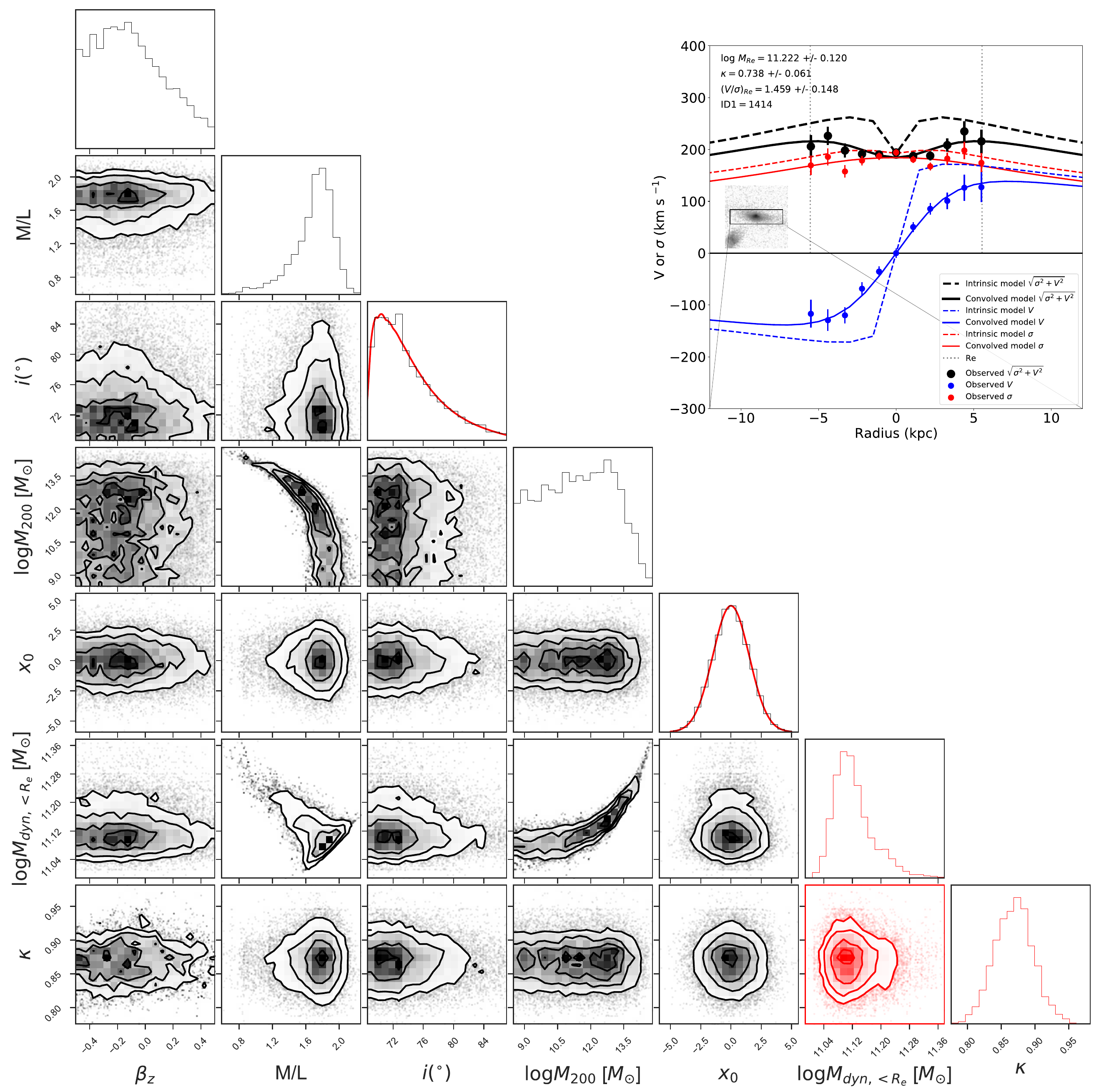} 
\caption{Covariance of the parameters of the JAM model fit to the observed $v_{rms}$ of galaxy $1414$ (see Fig.~\ref{fig:mge_example} for the light profile and HST image), with the best fitting dynamical model shown in the top right corner. The priors on the slit center, $x_{0}$, and the inclination, $i$, are shown in their respective one dimensional histogram as red lines. Contours indicate the 68\%, 90\% and 99\% confidence intervals.  There are clear degeneracies, most notably between stellar mass and dark matter mass. Some parameters (inclination and slit position) are not constrained at all by the data; marginalizing over these parameters helps to produce realistic uncertainties on the well-constrained physical parameters of interest:  $M_{dyn,<Re}$ and $\kappa$. These are shown in red because they are derived from fitting parameters, but not fitting parameters themselves.   The top-right panel shows the `convolved model' quantities include aperture and projection effects as well as seeing. The `intrinsic model' only include aperture and projection effects. The red and blue data points and models represent the line-of-sight velocity $V$ and velocity dispersion $\sigma$, respectively, and the black data points and models represent $v_{\rm{rms}}$. The spectroscopic aperture is show in the inset HST images. The quantities printed in the top-left corner of the panel refer to quantities presented in Sections \ref{label:JAM} and \ref{sec:catalog}. \label{fig:rms_examples}}
\end{figure*}

\section{Fitting Procedure}
\label{section:3.6}
\label{section:Section 4}

Following the methods and assumptions described in Sec.~3 we start with a fit of the line-of-sight second velocity moment of the JAM to the observed $v_{\rm{rms}} = \sqrt{v^{2}+\sigma^{2}}$. For axisymmetric galaxies that are centred in the slit the velocity profile is symmetric. Observing such a galaxy, one would expect observations at $+r$ and $-r$ to be independent observations of the same value. This is not true if the slit is slightly offset (see also Section \ref{subsection:effective seeing}) which is why we do not symmetrise the data. However, to ensure that asymmetric data points do not influence the fits too much, the uncertainties are set to the maximum of the formally measured uncertainties in $v_{\rm{rms}}$ and half the difference between the $v_{\rm{rms}}$ values at $+r$ and $-r$. In addition, at all radii a minimum uncertainty of $5$ km\,s$^{-1}$ is imposed to avoid that data points with spuriously low uncertainties dominate the fits. The fits extend to where we have \spel{measurements of} $\sigma$, even though the velocity measurements tend to extend further out. Finally, we measure and subtract the systematic velocity as a function of the slit center $x_{0}$ by fitting an isotropic ($\beta_{z}=0$), stellar-mass only model at every value of $x_{0}$.

\begin{deluxetable*}{ll}
\tablecaption{Free parameters in the fit, with assumed priors}
\tablehead{Parameter & Prior}
\startdata
$\beta$              & Flat prior on -0.5\textless{}$\beta$\textless{}0.5                  \\
M/L                  & Unconstrained                                                 \\
$v_{DM}$             & Flat prior on -600\textless{}$v_{dm}$\textless{}600 km$\cdot$s$^{-1}$             \\
Inclination          & $q$-dependent prior, assuming an intrinsic thickness distribution of $\mathcal{N}(0.41, 0.18)$ (see section $3.5$)               \\
Slit center, $x_{0}$ & Gaussian, centered on 0 with a $1.5$ pixel standard deviation
\enddata
\end{deluxetable*}

The fit is done using the Markov Chain Monte Carlo (MCMC) software \emph{emcee}\footnote{The EMCEE code can we downloaded from \url{https://emcee.readthedocs.io/en/latest/}} (\citealt{EMCEE2013}), which samples the parameters space with an ensemble of `walkers'. The five free parameters in the fit are the stellar mass-to-light ratio ($M/L$), the virial mass of the dark matter halo parameterized by the circular velocity ($v_c$), the inclination ($i$), the anisotropy $\beta_{z}$ and the slit centering $x_{0}$. We use 40 walkers, with a burn-in run of 80 iterations. Convergence was verified by visually examining the parameter values as the fit progresses, which usually occurs after several hundred iterations.

\begin{figure*}
\epsscale{.38}
\plotone{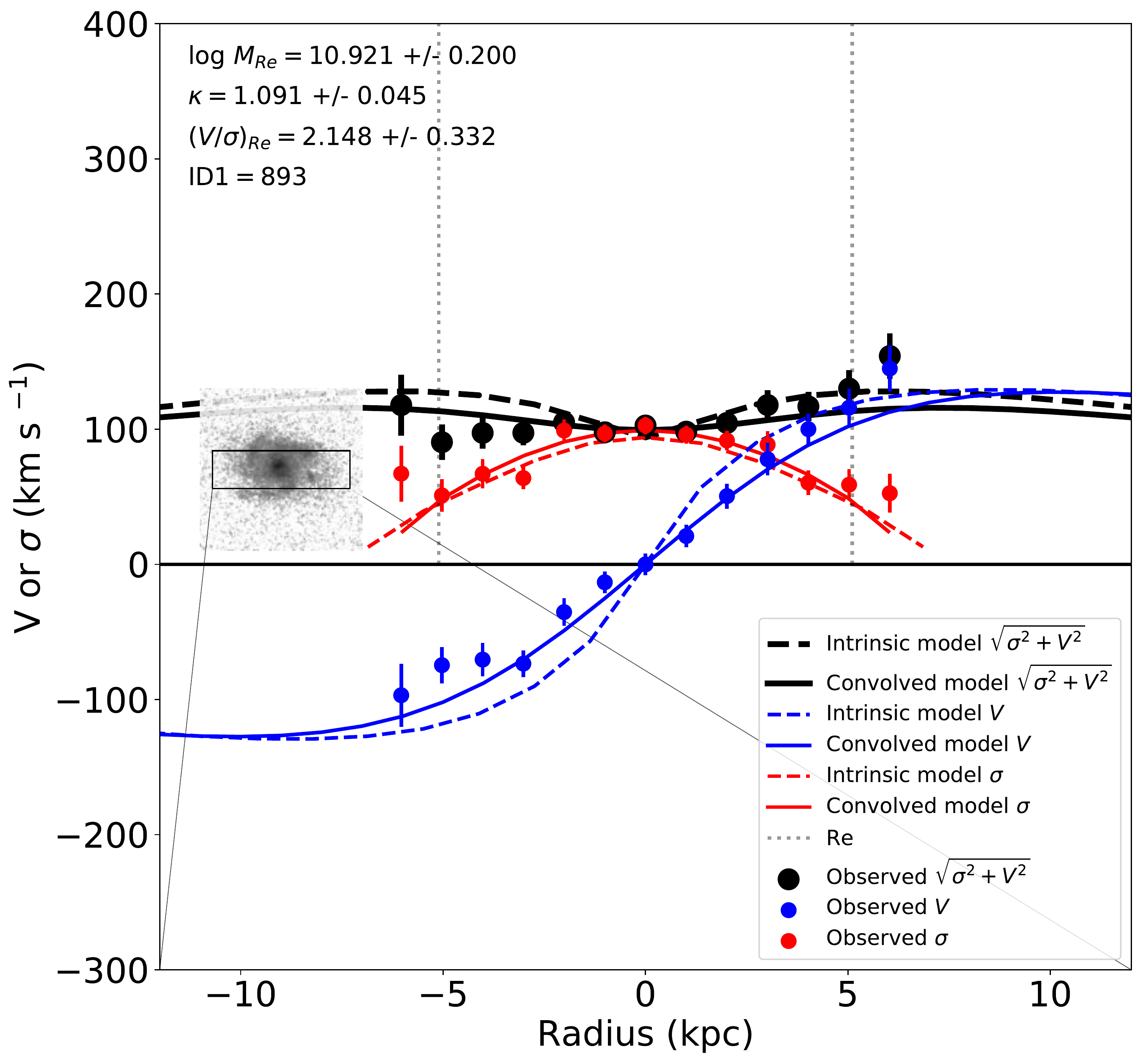}
\plotone{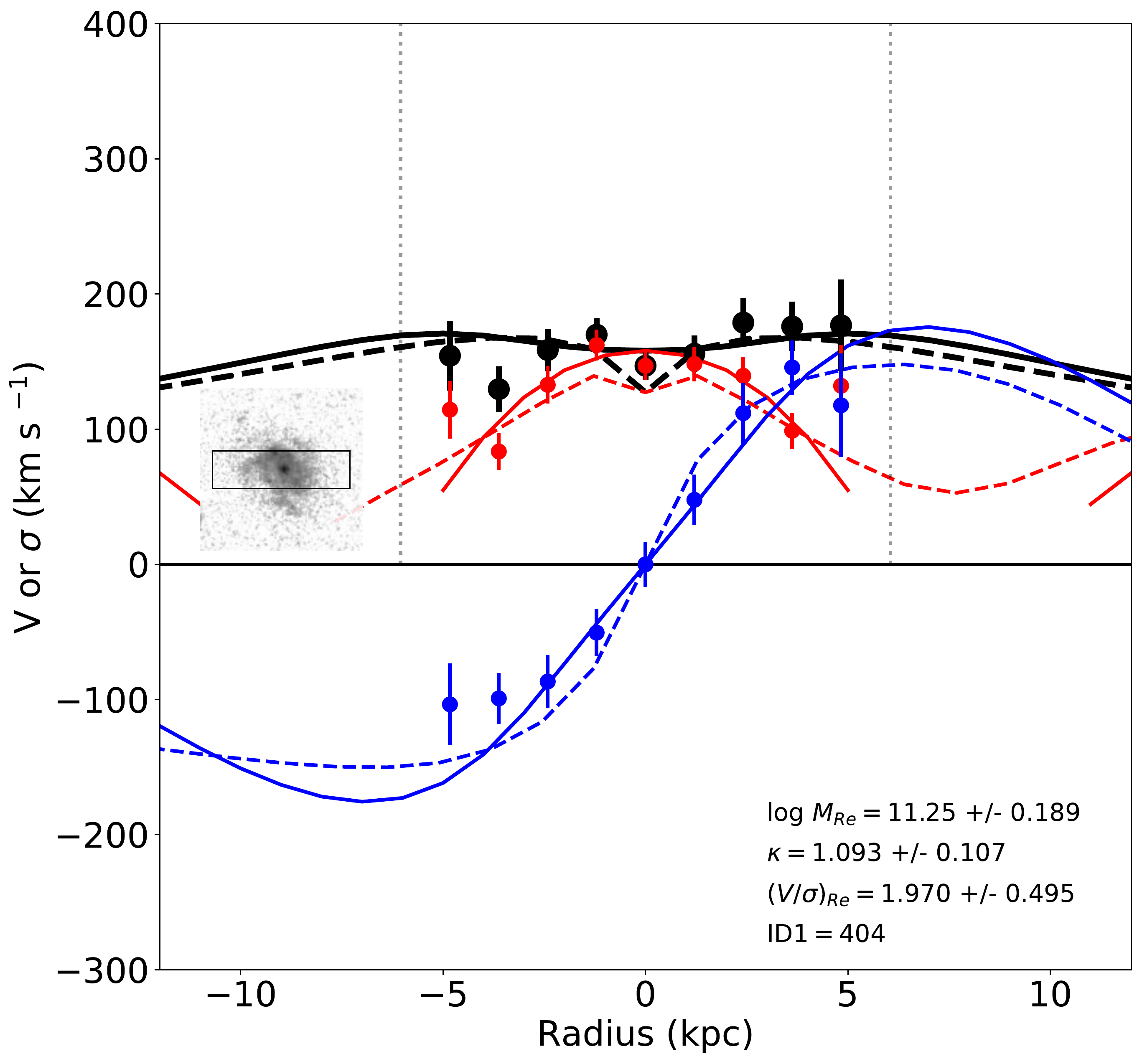}
\plotone{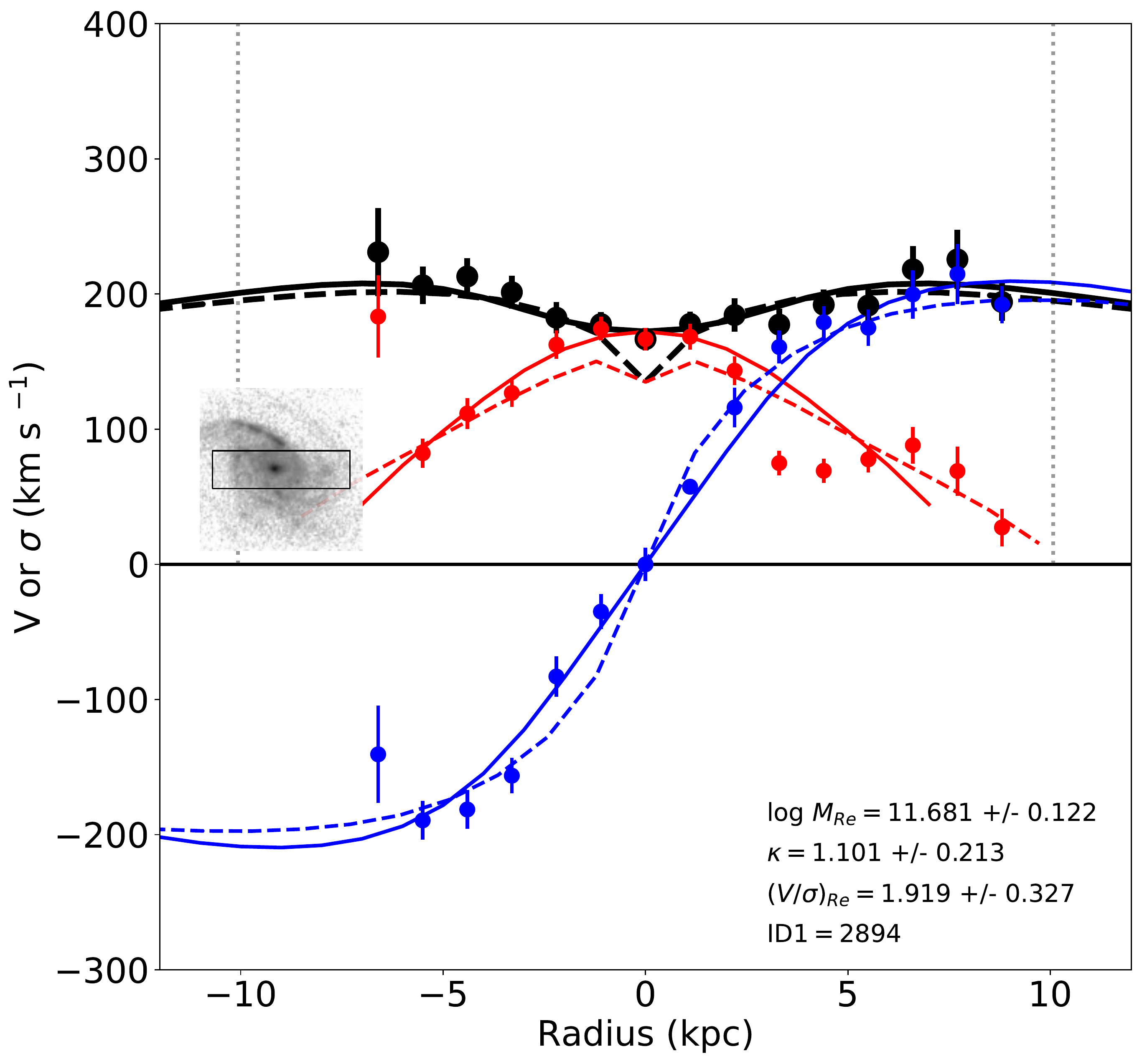}
\plotone{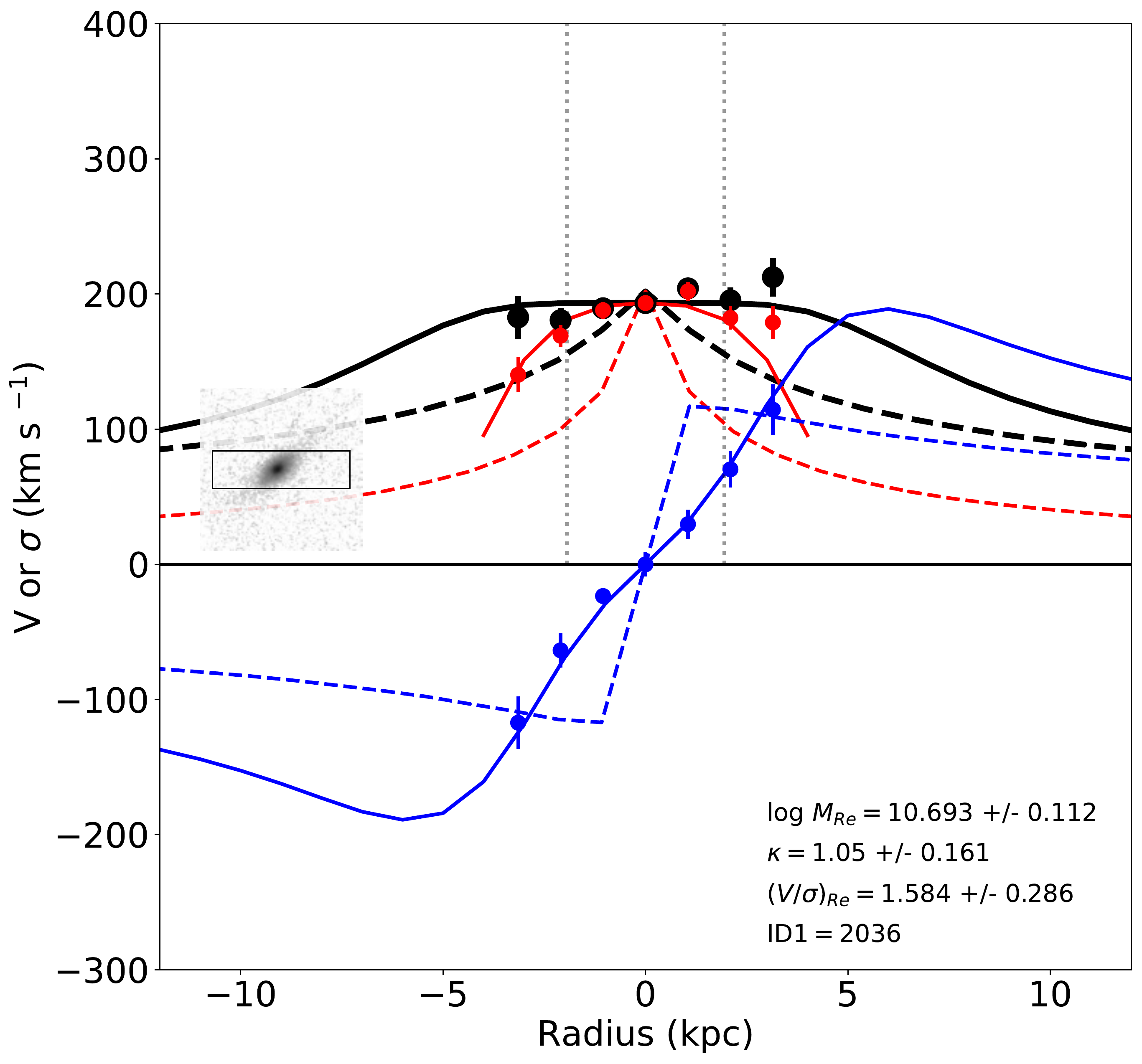}
\plotone{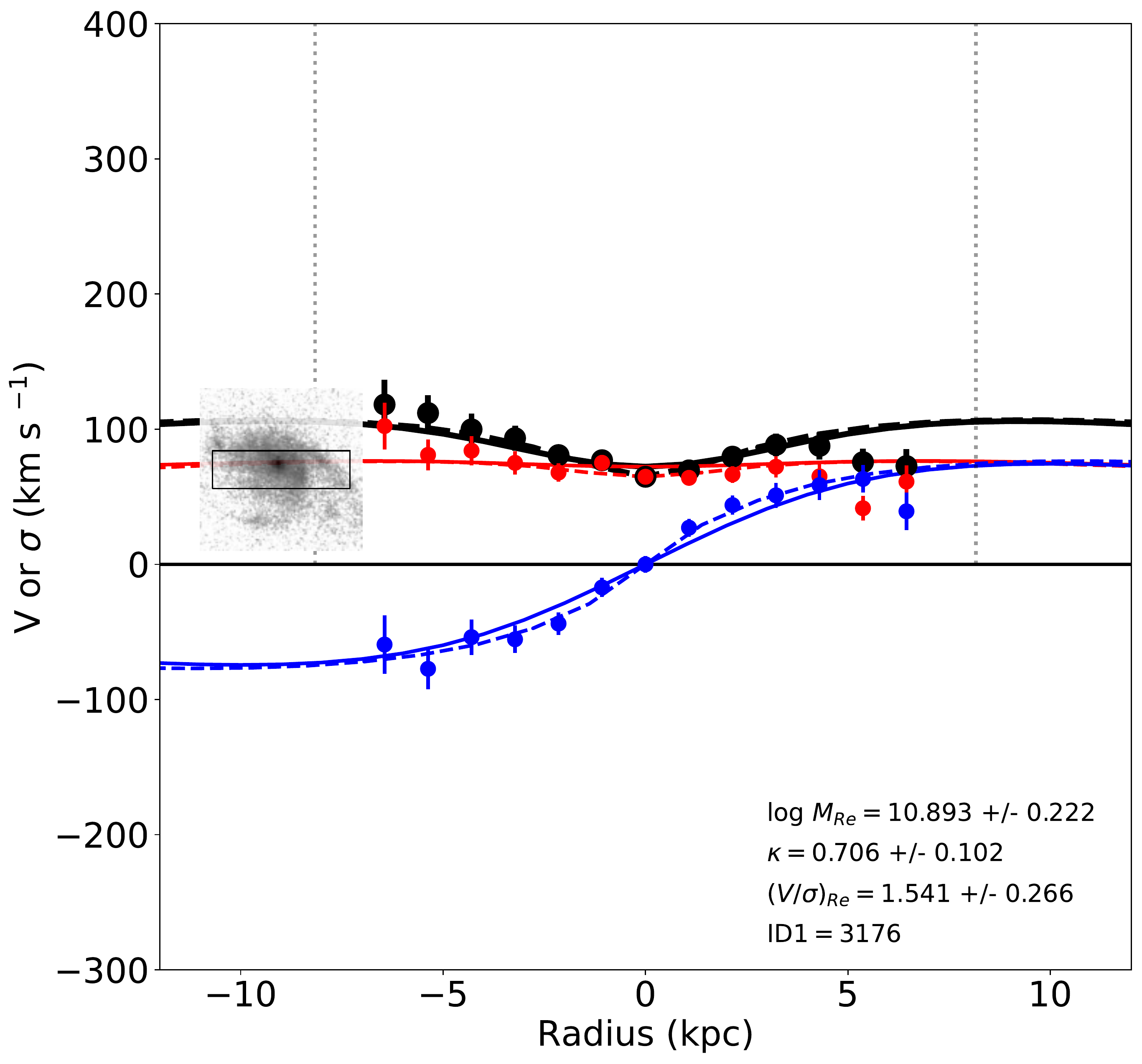}
\plotone{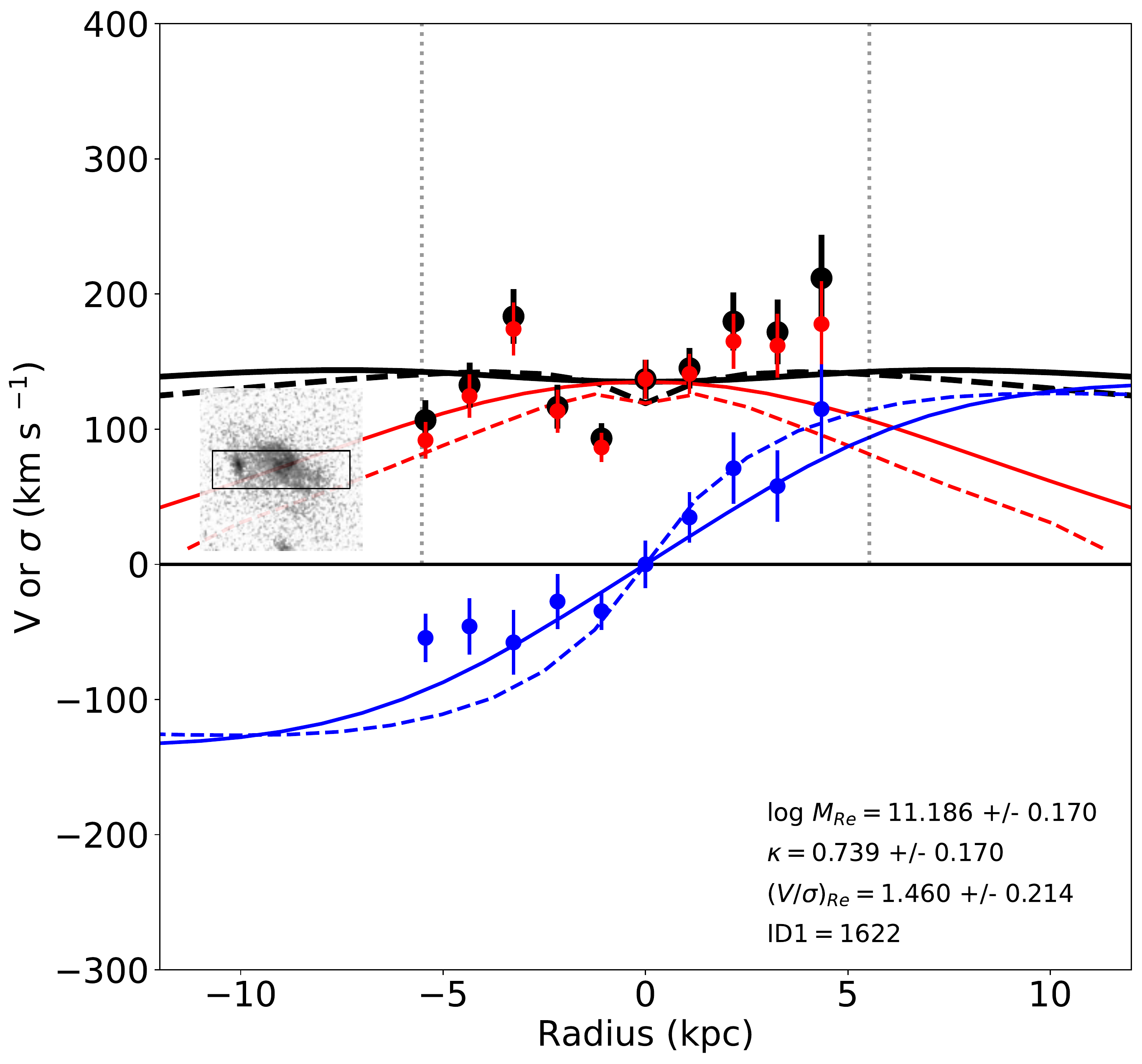}
\plotone{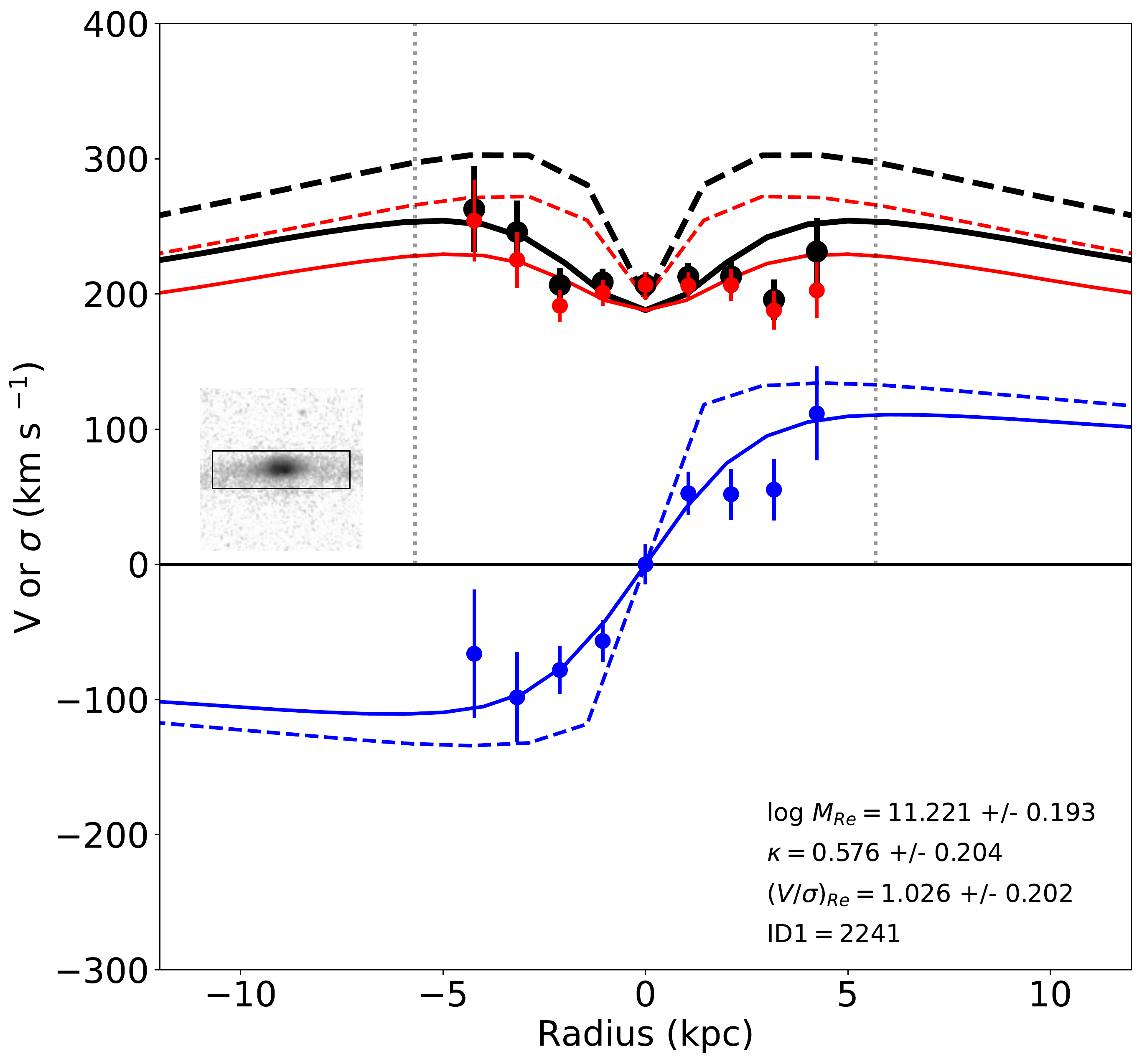}
\plotone{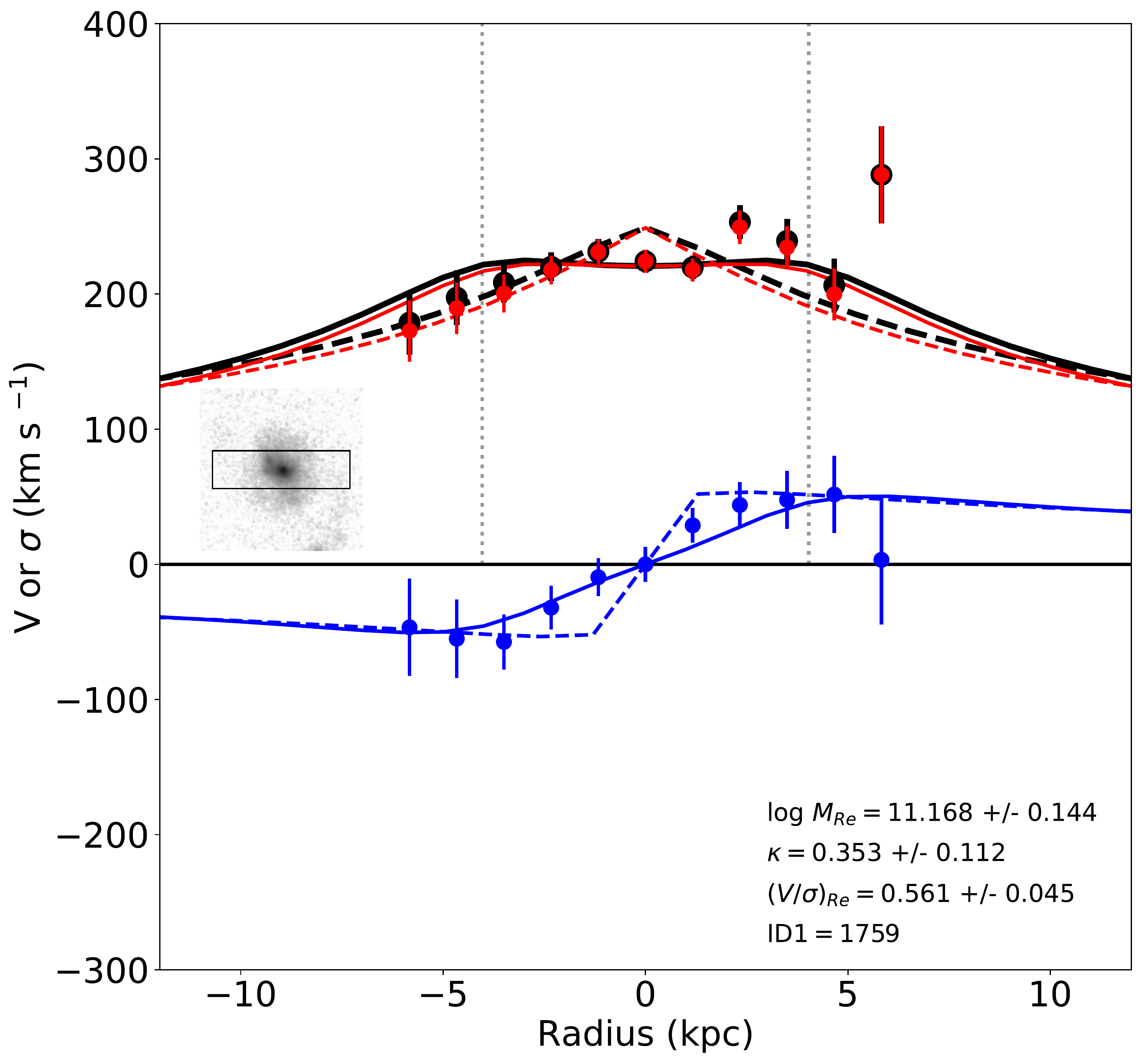}
\plotone{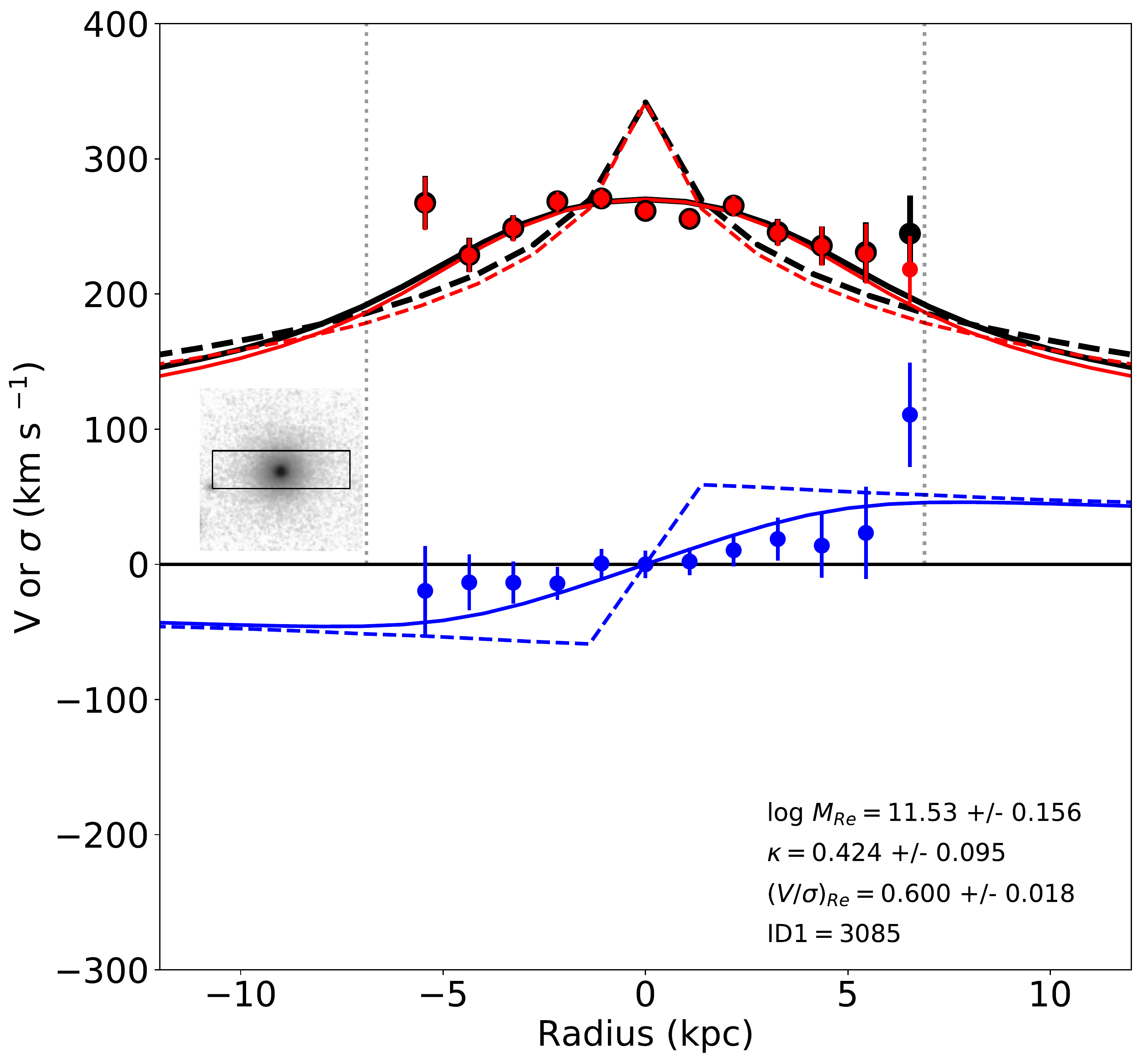}
\plotone{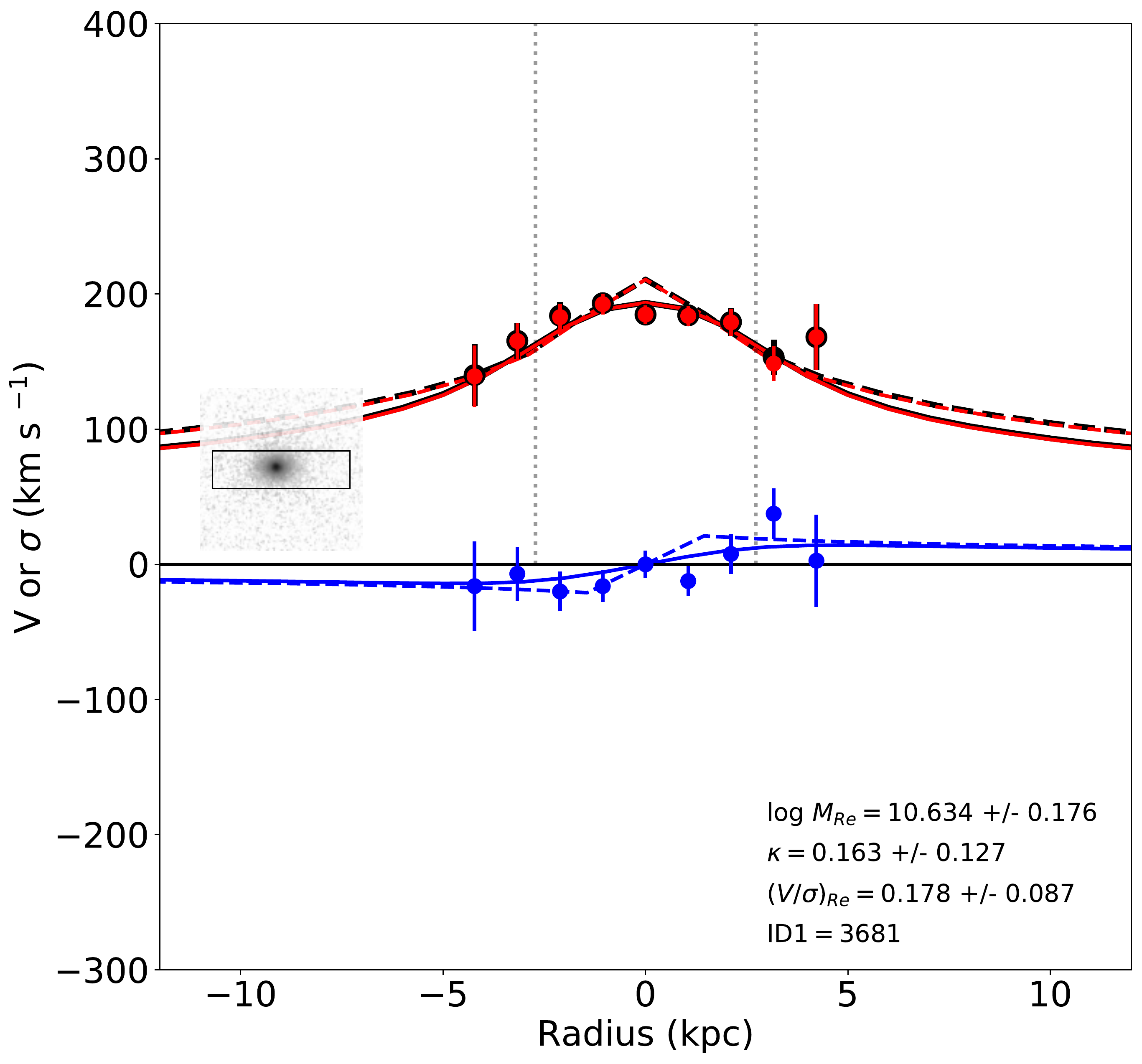}
\plotone{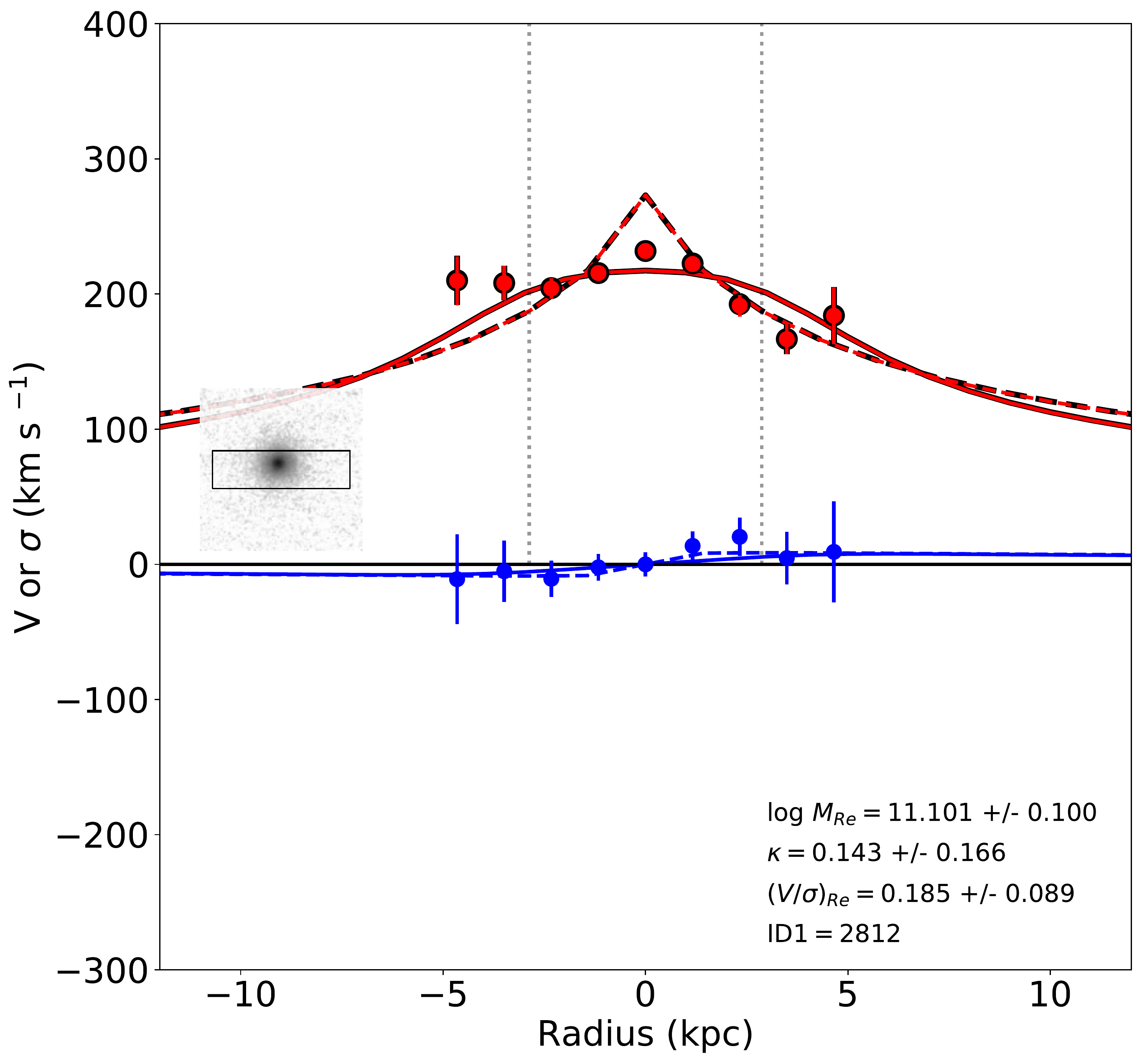}
\plotone{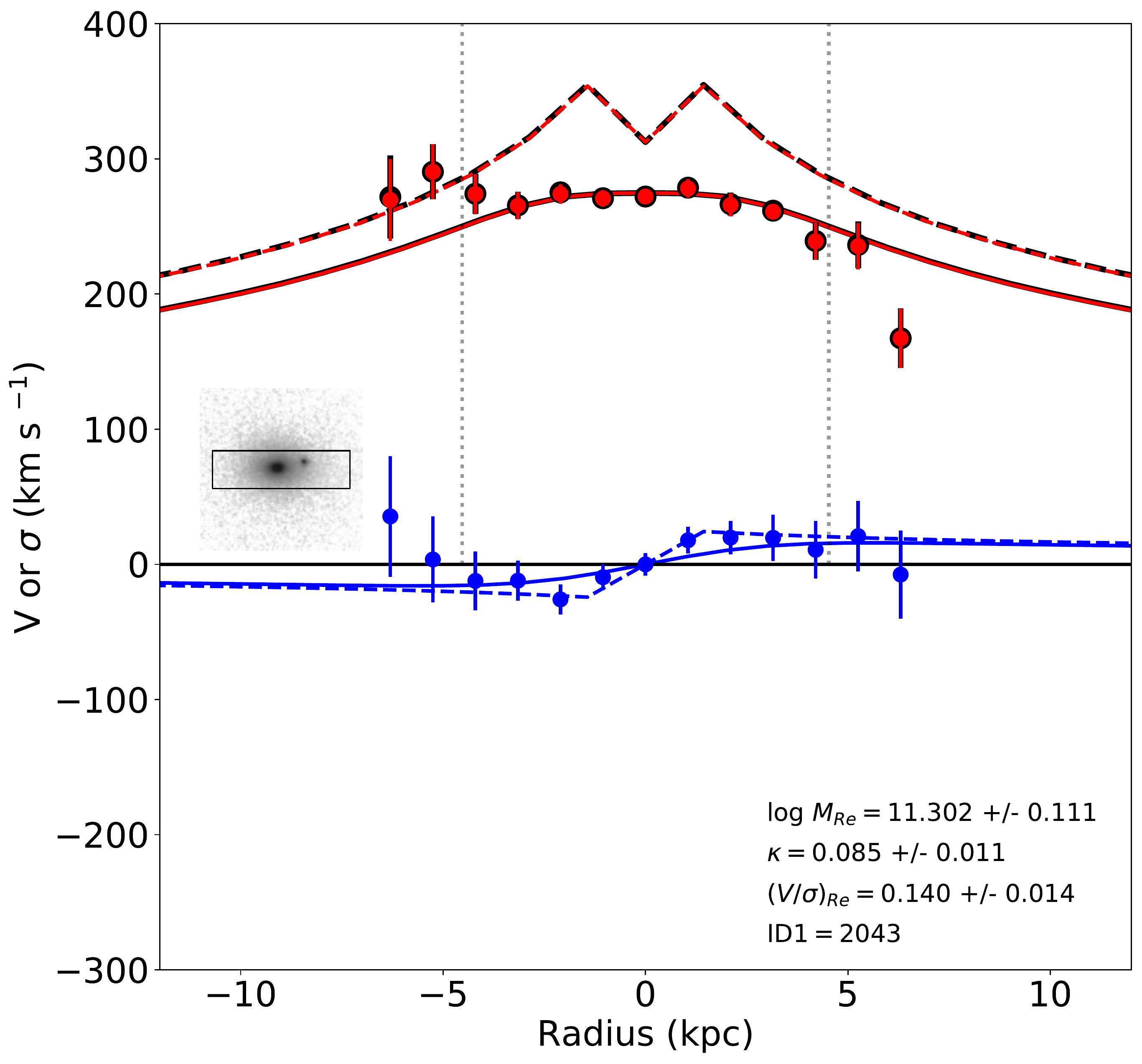}
\caption{Observed and best-fit model kinematic profiles for a subset of 12 galaxies.  The galaxies are ordered by mass (from left to right) and degree of rotational support (from bottom to top). The quantities printed in the top-left corner of each panel refer to quantities presented in Sections \ref{label:JAM} and \ref{sec:catalog}. The top-left panel includes a legend of the data points and models.  The `convolved model' quantities include aperture and projection effects as well as seeing. The `intrinsic model' only include aperture and projection effects. The red and blue data points and models represent the line-of-sight velocity $V$ and velocity dispersion $\sigma$, respectively, and the black data points and models represent $v_{\rm{rms}}$, the quadratic combination of $V$ and $\sigma$. The spectrosopic aperture is show in the inset HST images. \label{fig:velfits_page1}}
\end{figure*}

As an example, Figure \ref{fig:rms_examples} shows the covariance between the various parameters, for a galaxy that was chosen to be typical in terms of the observed velocity dispersion (192 km s$^{-1}$) and radial extent of the observed kinematics (5 pixels or 1$\arcsec$ -- one-sided).  The inclination is unconstrained by the data and is equal to the prior, as is the slit center $x_{0}$. $\beta_{z}$ is weakly constrained, as expected. Dark matter mass and stellar mass are degenerate, but the total mass within $R_{e}$ ($\log M_{dyn,<Re}$) is well constrained. Note that this is not an independent parameter but computed from $M_{200}$ and $M/L$. The best-fit models have a median reduced $\chi^{2}$ of $1.03$, so given the simplicity of the models the fits are more than adequate.

The posterior for the rotation parameter $\kappa$ is shown in Figure \ref{fig:rms_examples}, but is not fit simultaneously (see Section \ref{label:JAM}). Instead, the best-fit $\kappa$ is computed at every step of each Markov chain $(\beta_{z}$, $M/L$, $i$, $M_{200}$, $x_{0}$), without influencing the overall likelihood. The rotation parameter $\kappa$ has been used in the past as a computationally convenient way of disentangling rotational from random motion, and for an isotropic oblate rotator, $\kappa=1$ by definition. As long as the rotation is not larger than $v_{rms}$, $\kappa$ can be arbitrarily higher than $1$, which means that the model contains more rotation than is required to explain the observed flattening, and significant anisotropy (in the vertical direction) contributes to the thickness of the system. A maximum $\kappa=\kappa_{max}$ can be defined for which $v_{rms}=v_{rot}$. 
If the best-fit is found for $\kappa > \kappa_{max}$, we set $\kappa=\kappa_{max}$. These cases can occur for various reasons: if the galaxy has an anisotopy that differs from Eq.~\ref{eq:k}; if the assumptions of the axisymmetric modelling are violated; if the assumption of a spatially constant $\kappa$ is a poor one; or if $\kappa$ is an improper (or too simplistic) parameterisation in the first place.

Examples of the observed kinematic profiles and corresponding best-fit models in Figure \ref{fig:velfits_page1} are shown for 12 galaxies spanning the full dynamic range of mass and kinematic structure. The axi-symmetric models are sufficiently flexible to reproduce the full range of observed kinematic and structural properties: non-rotating and rotating galaxies, dispersion profiles that peak and dip in the center, late- and early-type galaxies, face-on and edge-on galaxies, aligned and mis-aligned galaxies, and compact and extended galaxies.

\section{Tests and Model Validation}
\label{section:Tests}

\subsection{Is a dark component required?}
\begin{figure*}
\epsscale{.5}
\plotone{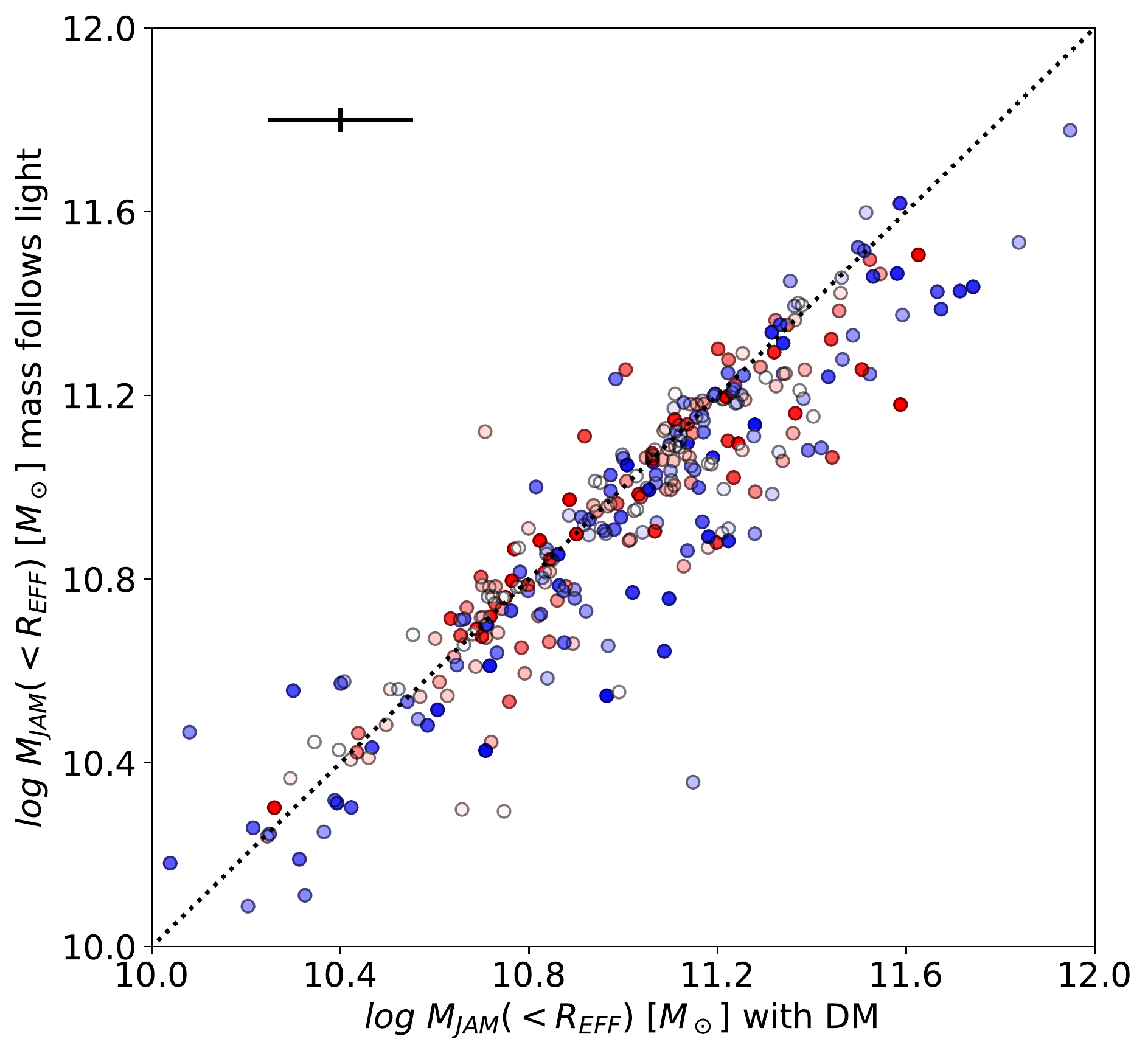}
\epsscale{.57}
\plotone{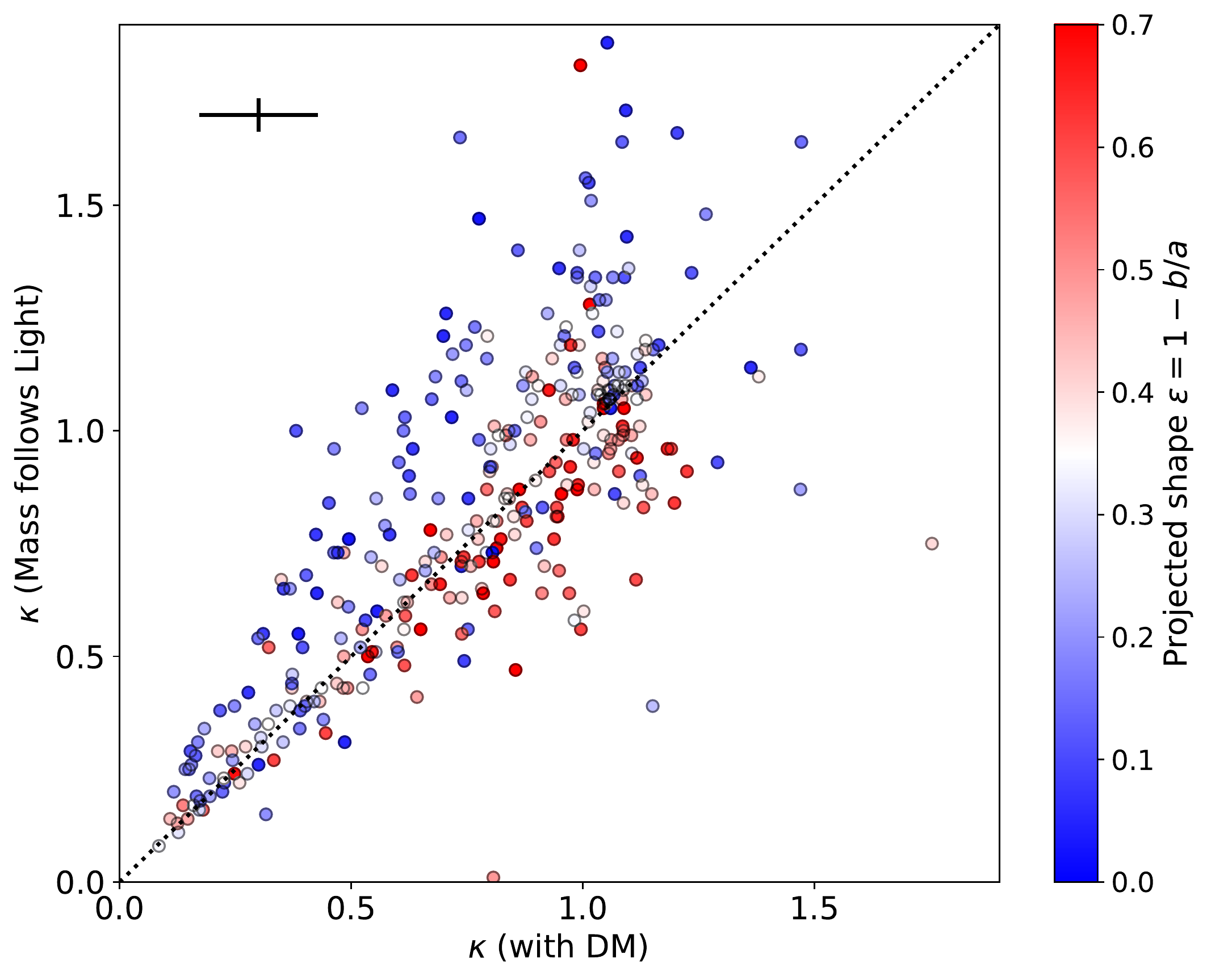} 
\caption{Model comparison for two different assumptions: with dark matter (our default model) and without dark matter (mass follows light). \textit{Left:} Enclosed mass within the effective radius; \textit{Right:} rotation parameter $\kappa$, both color coded with the projected ellipticity. Typical error bars are shown in the top-left corners.  For the majority of galaxies the mass estimates agree between the two models, but there is a prominent tail for which the mass-follows-light models produce lower mass estimates. Notably, the uncertainties for the mass follows light models are unrealistically small ($\sim6\%$), presumably due to the lack of flexibility in the model. The clear ellipticity dependence for $\kappa$ betrays strong projection effects. The mass-follows-light models produce systematically higher $\kappa$ for near face-on, rotating galaxies, which implies that those are overestimated and that mass-follows-light models do not accurately capture the dynamical structure of galaxies in a manner that is independent of viewing angle.
 \label{fig:misl_comparison}}
\end{figure*}

If we leave out the dark component, our model reverts to the commonly used mass-follows-light approach, where the total mass profile is identical to the MGE stellar light model, with only the $M/L$ scaling as a free parameter. We show the comparison in Figure \ref{fig:misl_comparison}. Importantly, the formal uncertainties are very small for the mass-follows-light model (typically, just 6\%). This is not realistic for total mass estimates of high-redshift galaxies with limited spatial resolution. The uncertainties for the default model are much larger and we verify in Section \ref{sec:duplicates} that these are realistic based on duplicate observations. The median differences in mass are modest ($<0.10$~dex) but for individual galaxies the default model often produces a mass that is two times higher than the mass-follows-light mass.  Moreover, there are strong and systematic differences in the $\kappa$ parameter that traces the degree of rotational support. For the mass-follow-light models the round (near face-on) galaxies have the highest $\kappa$ values.  This is counterintuitive and implies a bias: round galaxies should have similar or lower $\kappa$ values compared to flat galaxies.

On the basis of the biased $\kappa$ estimates and the unrealistically small uncertainties for the mass-follows-light models we conclude that the dark component is an essential ingredient. It allows for the required flexibility in modeling the observed kinematic data, especially for face-on disks. As discussed in Section \ref{sec:dark} the dark component does not necessarily reflect dark matter alone, but accounts for deviations from the mass follows light assumption due to stellar $M/L$ variations and the contribution from gas as well.

\subsection{Duplicate Observations}\label{sec:duplicates}
\begin{figure*}
\epsscale{1}
\plotone{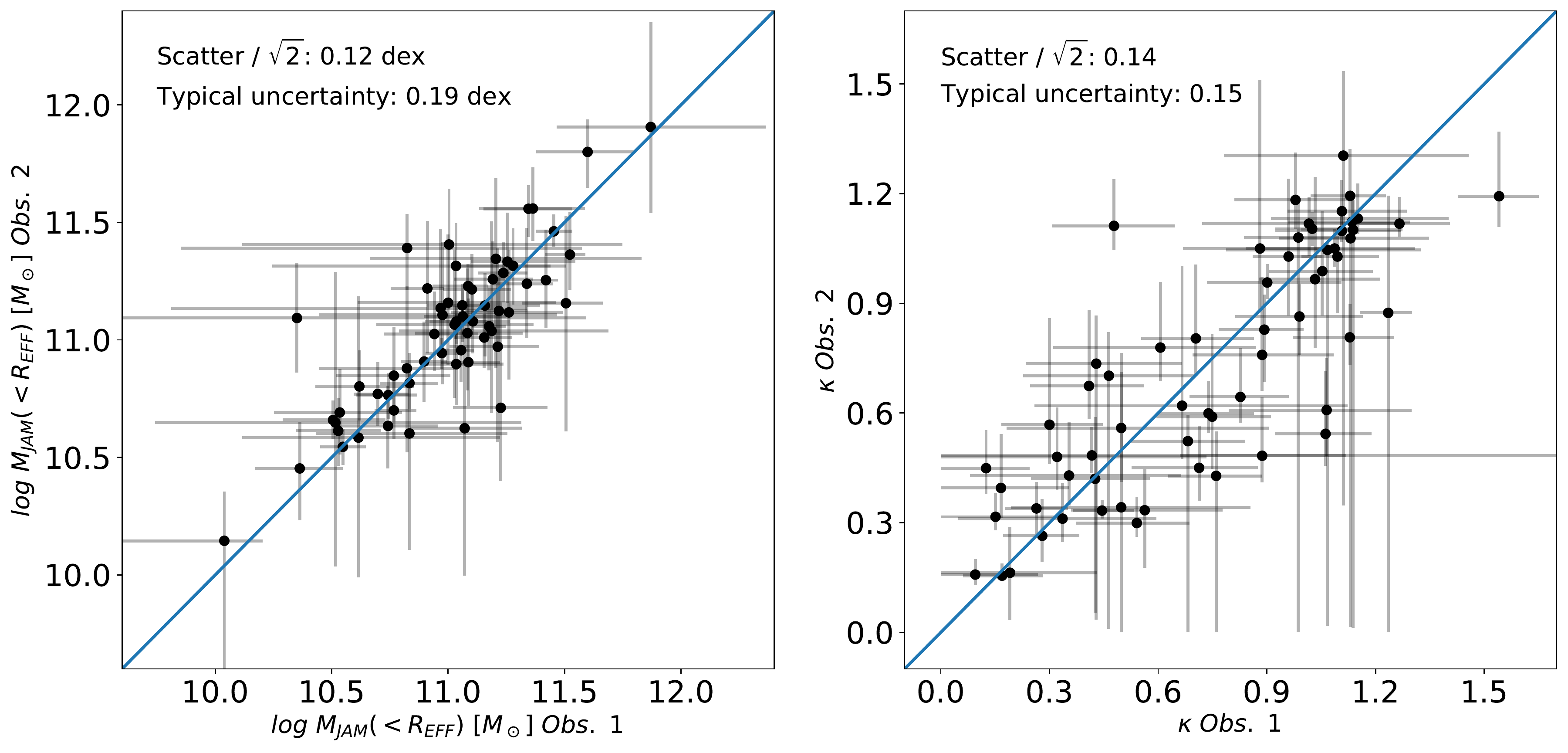}
\caption{Mass (left) and rotation parameter $\kappa$ comparison for galaxies with repeat observations. The scatter compares reasonably well with the formal uncertainties, and imply that the these uncertainties that are inferred from the posterior distributions are accurate. See text for details.\label{fig:duplicates}}
\end{figure*}

A number of galaxies have been observed multiple times in LEGA-C as a consequence of the observing approach. These provide independent kinematic measurements of the same object, so we can leverage this to test the robustness of the parameters we derive and obtain an independent estimate of all uncertainties associated with the spectroscopic data. In the sample with dynamical models there are 65 such duplicates. Figure \ref{fig:duplicates} compares the dynamical mass and rotation parameter $\kappa$ for these galaxies. The two sets of observations scatter around the one-to-one line with a scatter that corresponds reasonably well with the expectation from the formal (random) uncertainties. Theses uncertainties, inferred from the posterior distributions, are on average 50\% larger than the scatter for the mass estimate.  We attribute this to the flexibility in our model afforded by a mass component that does not follow the light distribution. The comparison of duplicate observations may imply that this conservative approach leads to overestimated random uncertainties.  On the other hand, the light tracer, inclination prior and other elements in the model are identical for the duplicates, and any uncertainties associated with those remain hidden in this test: the duplicate results are only independent in terms of kinematic measurements, but the light tracer, stellar mass model, inclination and anisotropy priors are identical. The prudent decision, then, is to use the uncertainties as inferred from the posterior distributions.

\section{Catalog Contents}\label{sec:catalog}
The dynamical modeling results are made available electronically, and Table \ref{tab:cat} shows an excerpt. We provide the total mass in 5 different spherical apertures: 1 kpc, 5kpc, 10kpc, \re\footnote{\re~is the semi-major axis of the ellipse that contains half of the S\'ersic model light.}, and 2\re; these are spherical apertures, rather than projected, cylindrical apertures.  A value is only included in the table if that aperture does not amount to a significant extrapolation given the radial extent of the stellar kinematic data. We allow for a maximum of 20\% extrapolation in radial extent. Given this constraint, the number of objects with mass estimates within these 5 apertures are 861, 809, 206, 730, and 319, respectively. As one would expect, for most objects the radial extend of the kinematic measurements is similar to the effective radius: the median kinematic radial extent-to-effective radius ratio is 1.45 (see Figure \ref{fig:rmax}).   

Please note the presence of a small subset of galaxies with negative mass estimates. This arises due to the prior on the dark mass component that is centered at zero to avoid a positive bias in the total mass estimates (see Section \ref{sec:dark}). All negative mass estimates have large uncertainties, so that a positive mass is never ruled out with more than 95\% confidence, as is expected given the derived uncertainties.  

The rotation parameter $\kappa$ is described in Section 3.1. The rotational velocity \Ve~is calculated at \re~for the edge-on, deprojected model. The effective velocity dispersion \se~is calculated for the light-weighted, face-on deprojected model within \re. The catalog contains best values and lower and upper error bars. These are obtained from the posterior distributions, where the best value is the median and the upper (lower) error bars reflect the difference between the 84th (16th) percentile and the median.

A rotation proxy that is popular in the literature is the projected quantity $\lambda_{R}$, which is frequently used by local IFU studies (and first introduced for SAURON in \citealt{Emsellem2007}). Calculating $\lambda_{R}$ directly from our data is not useful given the large seeing convolution kernel and single kinematic axis, so instead we could infer it from the model.  But, as shown by \citet{harborne19, harborne20}, even surveys of relatively nearby galaxies suffer from significant seeing convolution and the results would not comparable, and it would be misleading to name such a quantity $\lambda_R$. Moreover, given the dependence on the model it makes little sense to leave in inclination-dependent effects -- hence our choice to calculate face- and edge-on projected quantities.  

A direct comparison with present-day galaxies is, at this point, not possible: the only way to infer evolution in the intrinsic dynamical structure of galaxies is create the equivalent dynamical models for present-day galaxies.

In our own previous work (\citealt{Bezanson2018}) we measured $(v_{5}/\sigma_{0})$ without correcting for seeing or slit losses -- instead we used IFU data from CALIFA (\citealt{CALIFA2012}), downgraded to resemble our high-redshift data, in order to infer redshift evolution. Our models now provide the intrinsic $(v_{5}/\sigma_{0})$ and we find the effect of downgrading the CALIFA data agrees very well with the change from intrinsic (model) $(v_{5}/\sigma_{0})$ and directly observed $(v_{5}/\sigma_{0})$.  This validates the conclusions drawn by \citet{Bezanson2018} that high-redshift quiescent galaxies show more rotation than their present-day counterparts, which we interpreted as direct evidence for merging driving galaxy growth and angular momentum reduction.\\

\begin{deluxetable*}{cccccccccccc}
\tabletypesize{\scriptsize}
\tablecolumns{12}
\tablecaption{Stellar Dynamical Parameters \label{tab:cat}}
\tablehead{
\colhead{ID1} &  \colhead{ID2} &  \colhead{$r_{\rm{max}}$} &  \colhead{$R_e$} & \colhead{$\log M_{1}$} & \colhead{$\log M_{5}$} &  \colhead{$\log M_{10}$}  & \colhead{$\log M_{R_e}$}  & \colhead{$\log M_{2R_e}$}  & \colhead{$\kappa$} & \colhead{$V_{R_e}$} & \colhead{$\sigma_{R_e}$} \\
\colhead{} & \colhead{} & \colhead{kpc} & \colhead{kpc} & \colhead{$M_\odot$}  & \colhead{$M_\odot$} & \colhead{$M_\odot$} & \colhead{$M_\odot$} & \colhead{$M_\odot$} & \colhead{} & \colhead{$km~s^{-1}$} & \colhead{$km~s^{-1}$}
}
\startdata
5 & 4792 & 6.9 & 5.0 & $9.99^{+0.08}_{-0.06}$ & $10.88^{+0.27}_{-0.24}$ & \nodata & $10.88^{+0.27}_{-0.24}$ & \nodata & $1.10^{+0.07}_{-0.04}$  & $179^{+27}_{-34}$ & $90^{+5}_{-4}$ \\
15 & 6556 & 4.7 & 3.2 & $10.68^{+0.03}_{-0.03}$ & $10.73^{+0.53}_{-0.35}$ & \nodata & $10.84^{+0.18}_{-0.12}$ & \nodata & $0.34^{+0.20}_{-0.34}$  & $69^{+7}_{-90}$ & $204^{+13}_{-16}$ \\
16 & 6859 & 4.9 & 7.6 & $10.83^{+0.04}_{-0.03}$ & $11.33^{+0.17}_{-0.16}$ & \nodata & \nodata & \nodata & $0.36^{+0.15}_{-0.06}$ &  $120^{+4}_{-2}$ & $214^{+9}_{-11}$ \\
19 & 7419 & 13.0 & 11.5 & $9.72^{+0.25}_{-0.31}$ & $11.00^{+0.19}_{-0.20}$ & $11.46^{+0.21}_{-0.20}$ & $11.54^{+0.23}_{-0.21}$ & \nodata & $0.78^{+0.18}_{-0.09}$ &  $123^{+7}_{-7}$ & $160^{+13}_{-13}$ \\
26 & 10462 & 4.4 & 3.9 & $9.80^{+0.14}_{-0.12}$ & $10.70^{+0.52}_{-0.39}$ & \nodata & $10.55^{+0.46}_{-0.36}$ & \nodata & $0.57^{+4.42}_{-0.57}$  & $80^{+6}_{-19}$ & $58^{+10}_{-12}$ \\
27 & 10902 & 5.9 & 12.0 & $10.22^{+0.07}_{-0.05}$ & $11.02^{+0.25}_{-0.25}$ & \nodata & \nodata & \nodata & $1.04^{+3.96}_{-1.04}$ &  $211^{+34}_{-71}$ & $119^{+11}_{-10}$ \\
38 & 14375 & 6.2 & 2.8 & $10.76^{+0.02}_{-0.02}$ & $11.27^{+0.17}_{-0.16}$ & \nodata & $11.11^{+0.07}_{-0.08}$ & $11.29^{+0.19}_{-0.18}$ & $0.70^{+0.10}_{-0.06}$  & $241^{+5}_{-6}$ & $208^{+14}_{-31}$ \\
39 & 14729 & 6.3 & 10.4 & $10.36^{+0.06}_{-0.06}$ & $11.21^{+0.16}_{-0.18}$ & \nodata & \nodata & \nodata & $1.18^{+0.30}_{-0.08}$ &  $306^{+43}_{-53}$ & $175^{+16}_{-17}$ \\
62 & 25994 & 4.7 & 3.1 & $10.41^{+0.05}_{-0.05}$ & $11.10^{+0.19}_{-0.23}$ & \nodata & $10.91^{+0.11}_{-0.16}$ & \nodata & $1.08^{+0.05}_{-0.12}$ &  $249^{+22}_{-49}$ & $145^{+7}_{-7}$ \\
68 & 26283 & 7.8 & 4.2 & $10.51^{+0.04}_{-0.03}$ & $10.95^{+0.33}_{-0.35}$ & \nodata & $10.91^{+0.26}_{-0.28}$ & $11.01^{+0.71}_{-0.73}$ & $0.13^{+0.04}_{-0.13}$ & $29^{+1}_{-34}$ & $157^{+7}_{-9}$ \\
\vdots & \vdots & \vdots & \vdots & \vdots & \vdots & \vdots & \vdots & \vdots & \vdots & \vdots
\enddata
\tablecomments{(1): LEGA-C ID (DR3); (2):Ultra-VISTA ID \citep{Muzzin2013b}; (3): Extent of stellar kinematic data; (4): Effective radius of the galaxy; (5-9): Model mass enclosed in spherical apertures; (10): Rotation parameter; (11) Rotational velocity at $R_{e}$~(edge-on); (12): Velocity dispersion within $R_{e}$~(face-on). Values and uncertainties are based on 16th, 50th, 84th percentiles of posterior parameter distributions. The machine-readable table has 861 entries. We choose the following notation for a negative mass (see text): $\log M = -10$ means $M=-10^{10}$ (not $M=10^{-10}$).} 
\end{deluxetable*}

\section{Stellar dynamical structure of $z\sim0.8$ Galaxies}
\label{section:Section 5new}
\begin{figure*}[h!]
\epsscale{1.1}
\plotone{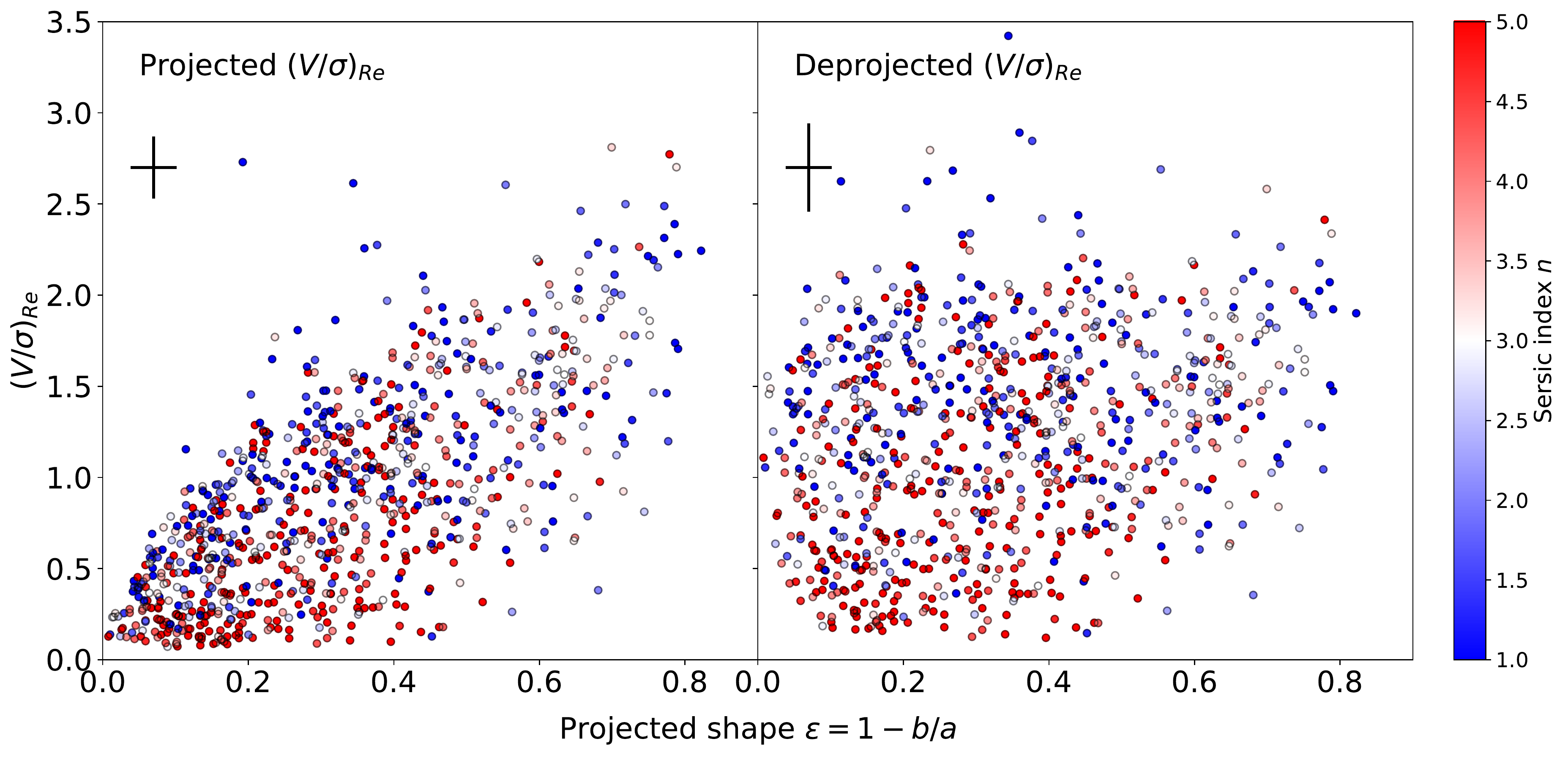} 
\caption{Projected (left) and deprojected (right) \Vsig vs.~projected shape, color coded with S\'ersic index $n$. Flat projected shapes represent a fairly good prediction for a high degree of rotational support, but for individual galaxies a round shape is a poor predictor. High-S\'ersic index galaxies show a large range in rotational support, overlapping with the distribution of low-S\'ersic index galaxies, but with an additional population of slowly rotating galaxies. \label{fig:e_vsig}}
\end{figure*}

\begin{figure*}[h!]
\epsscale{1.1}
\plotone{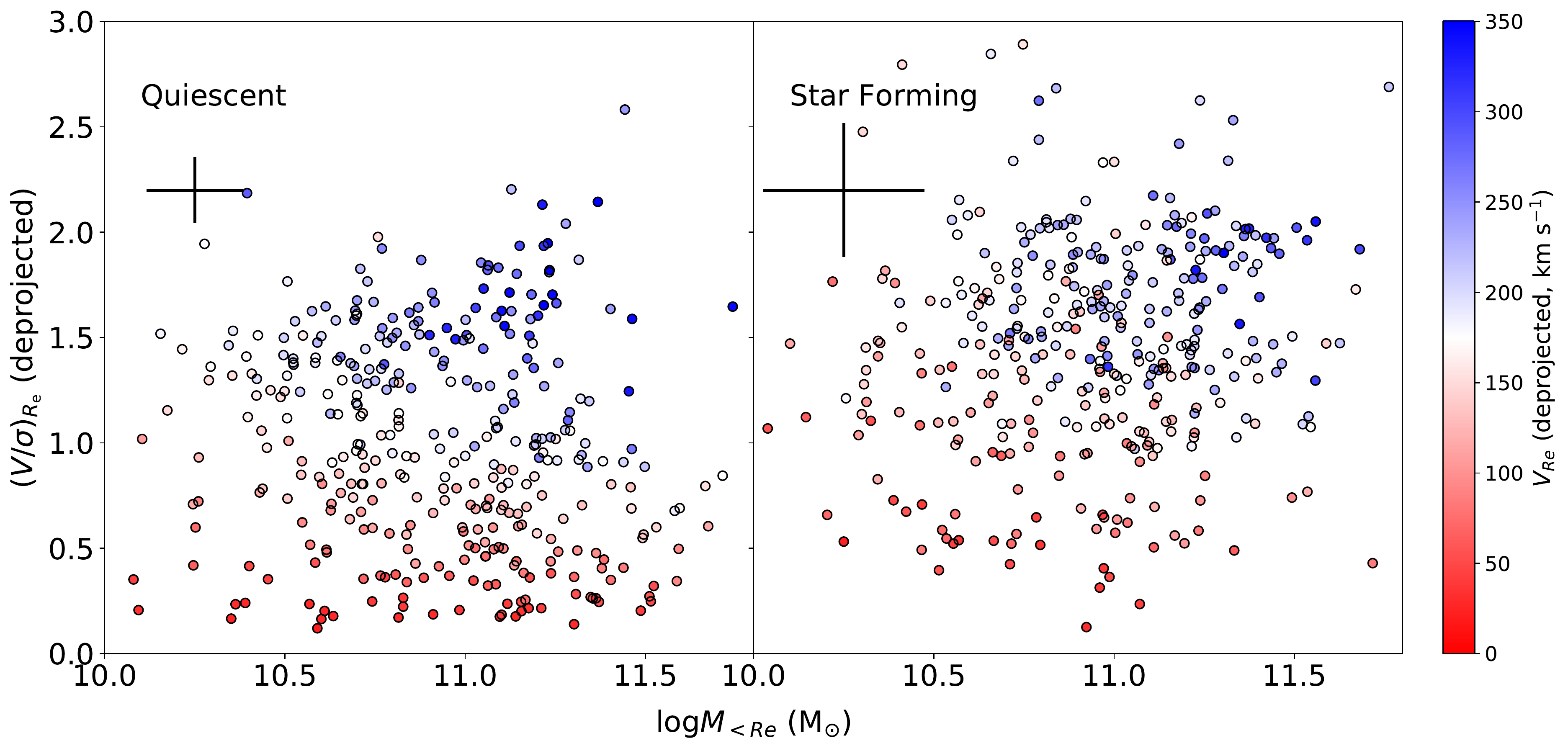} 
\caption{Deprojected \Vsig vs.~dynamical mass within $R_e$ for quiescent (right) and star-forming galaxies (right), color coded with deprojected (edge-on) rotation velocity $V$ at $R_e$. The correlations are generally weak, and quiescent galaxies overlap strongly with star-forming galaxies in terms of \Vsig. Only at the high-mass end there is a clear difference between the populations, where quiescent galaxies rotate slowly, and star-forming galaxies are strongly rotation-dominated.\label{fig:m_vsig}}
\end{figure*}

It is beyond the scope of this paper to explore rotational support, angular momentum and the masses of galaxies in great detail, and in forthcoming papers we will explore the connection between star-formation history and dynamical structure, the various mass components of galaxies, and the comparison with dynamical modeling based on ionized gas kinematics. But to illustrate the richness of our data we show in Figures \ref{fig:e_vsig} and \ref{fig:m_vsig} the distribution of galaxy shapes, \Vsig and dynamical mass $M_{R_e}$.

The left-hand panel of Figure \ref{fig:e_vsig} is strongly reminiscent of integral field surveys of present-day galaxies \citep[e.g.,][]{Emsellem2011, van-de-sande17, falcon-barroso19}, with a strong correlation between projected shape and (projected) degree of rotational support. Low S\'ersic index (late-type) galaxies generally have higher \Vsig than high-S\'ersic index (early-type) galaxies, even at fixed projected ellipticity. In particular, we see a population of rather flat ($\epsilon\sim0.4$), but non-rotating high-$n$ galaxies. Such galaxies also exist in the present-day Universe and can be either anisotropically flattened systems or dynamically complex galaxies with, for example, counter-rotating disks \citep{rix92}. The spatial resolution of our data is obviously insufficient to distinguish between the different explanations, but we can conclude that the net rotation is low compared to the degree of flattening. There also is a small set of data points at very large \Vsig, away from the general locus: here the best-fit dynamical model has a high anisotropy (with very small vertical dispersion), but with large uncertainties (not shown in the Figure, but included in the catalog). 

The crowding in the lower-left part of the figure (round, non-rotating galaxies) is largely due to projection effects. Therefore, in the right-hand panel we show the deprojected \Vsig (as explained in Section \ref{sec:catalog}). Even for nearly round galaxies with $\epsilon\sim0.1$ our modeling successfully recovered the degree of rotational support, as indicated by the lack of a correlation between \Vsig and $\epsilon$ for low-$n$ galaxies. This population must consist of galaxies with a fairly homogeneous dynamical structure, with \Vsig$\sim1-2$, and for which the scatter in projected \Vsig is mostly explained by viewing angle (inclination). The high-$n$ galaxies are less homogeneous in dynamical structure. Like in the present-day Universe, there is a population of intrinsically round, slowly-rotating galaxies, mixed with a population that is fast rotating, and similar in dynamical structure as the low-$n$ population. We note that our models successfully distinguish between rotation- and dispersion-dominated galaxies; our inclination prior, based on an intrinsically oblate shape, did not preclude this. 

It is well documented that S\'ersic index correlates well with star-forming activity at all redshifts \citep{bell12}, and it is informative to split the population into quiescent and star-forming galaxies on the basis of their rest-frame $U-V$ and $V-J$ colors, computed as explained in our 2nd data release paper \citep{straatman18}. In Figure \ref{fig:m_vsig} we show the \Vsig distribution as a function of dynamical mass $M_{R_e}$ for quiescent and star-forming galaxies. For the quiescent galaxies we see that the most massive galaxies, as in the present-day Universe, are the most likely to be slowly rotating, while for the star-forming population we see that the most massive galaxies are the most rotation dominated. This immediately implies different evolutionary pathways for the most massive galaxies, depending on their star-formation history. 

The trend for star-forming galaxies may at first sight seem counter-intuitive, given that the most massive star-forming galaxies are more bulge-dominated than low-mass star-forming galaxies (which we see for our sample by an increase in S\'ersic index). Yet these more bulge-dominated galaxies have higher \Vsig. But these trends are not mutually exclusive: indeed, $\sigma$ increases with mass, signifying the presence of more massive bulges, but $V$ increases faster with mass than $\sigma$, signifying the presence of a large, extended disk.  Interestingly, however, the galaxies with the highest rotation speeds are quiescent: these are compact objects with high velocity dispersions ($>200$~km s$^{-1}$), yet flattened and rotation dominated, with $V_{R_e}\sim350$~km s$^{-1}$, reminiscent of what has been referred to as `fossil' galaxies in the present-day Universe \citep[e.g.,][]{van-den-bosch12, ferre-mateu12}.

If we then consider the interesting set of low-mass star-forming galaxies with little rotation, visual inspection of the HST images and the rotation curves suggest that this can occur for several reasons: some of these galaxies are somewhat irregular systems for which the assumption of axisymmetry is likely invalid, but the assessment of low \Vsig~still accurate; some galaxies are truly non-rotating systems and appear to be young, spheroidal galaxies that just fall short of being classified as quiescent; for other objects the stellar light distribution obviously does not follow the stellar mass distribution, and bright, blue star-forming regions or spiral arms affect the analysis of the HST light profiles, and thereby the geometry of the adopted axisymmetric system in the dynamical model.

\section{Summary \& Conclusions}
\label{section:Section 5}
We present axisymmetric Jeans models for 797 z$=0.6-1$ galaxies with spatially resolved stellar kinematic measurements from the LEGA-C survey (\citealt{LEGAC2016,straatman18,van-der-wel21}). The $K_s$-band selected parent sample of over 3000 galaxies makes no selection on color or morphology, and we select the subset objects that satisfy additional criteria ($S/N$, slit alignment, and regular morphology), but is otherwise unbiased except for a bias against edge-on disks due to attenuation. This is by far the largest sample of galaxies with resolved stellar kinematics at this redshift to date, by a factor 30, and the only sample comprising star-forming / late-type galaxies, as previous samples were focused solely on passive galaxies. The application of Jeans models is important because unlike galaxies in the present-day Universe, high-redshift galaxies are at most a few arcseconds in extent - not much lager than both the width of the slits ($1$") and the ground-based seeing ($\sim0.8$").  Our primary concern is to correct the observed rotation and velocity dispersion profiles (\citealt{Bezanson2018}) for the convolution by slit and seeing, and an unknown inclination angle.

We use \emph{HST} $F814W$ imaging to describe the tracer light. The underlying gravitational potential is modelled with a mass-follows light stellar component plus a NFW dark matter halo. Other free parameters are the inclination, the velocity anisotropy, and the positioning of the galaxy within the slit; these three parameters are not well constrained by our data, but we assess their impact on rotation and mass estimates by marginalizing over well-informed prior distributions. We provide a full treatment of the seeing and slit geometry on a galaxy-by-galaxy basis, and the JAM methodology is adapted to accept non-square pixels (Section \ref{section:slit effects}). We derive (total) dynamical masses and proxies for the total, \emph{intrinsic} rotational support of galaxies from these models (Table \ref{tab:cat}).

This data product provides the community with the tools to study the masses and kinematic properties of a large sample of galaxies, which was not possible before at large look-back time. For the first time we show the \Vsig distribution of galaxies at large look-back time (Fig.~\ref{fig:e_vsig}) and the complexity in evolutionary pathways revealed by the large variety in dynamical structure seen even for galaxies with the same (dynamical) mass (Fig.~\ref{fig:m_vsig}). 

These results are timely because cosmological hydrodynamical simulations have now reached the resolution to predict the stellar dynamical structure for large, representative samples. Further analysis of the models and the related data products will improve our understanding of these galaxies even further, and this will be published in a number of follow-up projects in the near future.

\acknowledgments
This project has received funding from the European Research Council (ERC) under
the European Union’s Horizon 2020 research and innovation programme (grant agreement 683184). Based on
observations made with ESO Telescopes at the La Silla
or Paranal Observatories under programmes ID 194.A2005 and ID 1100.A-0949 (the LEGA-C Public Spectroscopic Survey). JvdS acknowledges support of an Australian Research Council Discovery Early Career Research Award (project number DE200100461) funded by the Australian Government.

\software{EAZY \citep{EAZY2008}, JAM \citep{JAM2008}, MGE \citep{MGE2002}, emcee \citep{EMCEE2013}, corner.py \citep{corner}}

\clearpage
\bibliography{sample}{}
\bibliographystyle{aasjournal}

\appendix
\restartappendixnumbering

\section{Slit Mis-Centering}
\label{AppendixB}

For most purposes, the slit can be assumed to be centered on the center of the galaxy. This is how the masks are designed, with accurate coordinates available from \emph{HST} astrometry. Furthermore, given the angular sizes of galaxies compared to the slit width, even with a $1$ pixel offset in either direction most light would still be captured.  Even so, the kinematics will be affected by an offset, leading to (small) asymmetries and offsets in the observed velocity and dispersion profiles.

Here we  ask the question whether it is possible to constrain the slit position by comparing the observed LEGA-C light profile with the expected light profile from a rebinned, convolved \emph{HST} image with a slit placed at different offsets. For single, symmetric sources there are too many degeneracies to make this feasible, but galaxies with asymmetric profiles, or galaxies with a close neighbour that happens to fall within the slit, make this a viable option (Figure 14). Owing to their relative position, the flux ratio between the two sources is a function of the assumed offset (perpendicular to the slit, in the wavelength direction), which constrains the observed offset.

Unfortunately, such sources are scarce, with on average $\sim5$ useful examples per mask. This is insufficient for a robust determination, as there is no guarantee that galaxies within the same mask or even the same quadrant will share the same offset. However, given the ensemble of offsets across the entire survey the majority suggest offsets of $1-2$ pixels around $0$, with incidental suggestions of bigger offsets, justifying the assumed prior in Section \ref{section:Section 3}.

\begin{figure}[htb!]
\epsscale{0.8}
\plotone{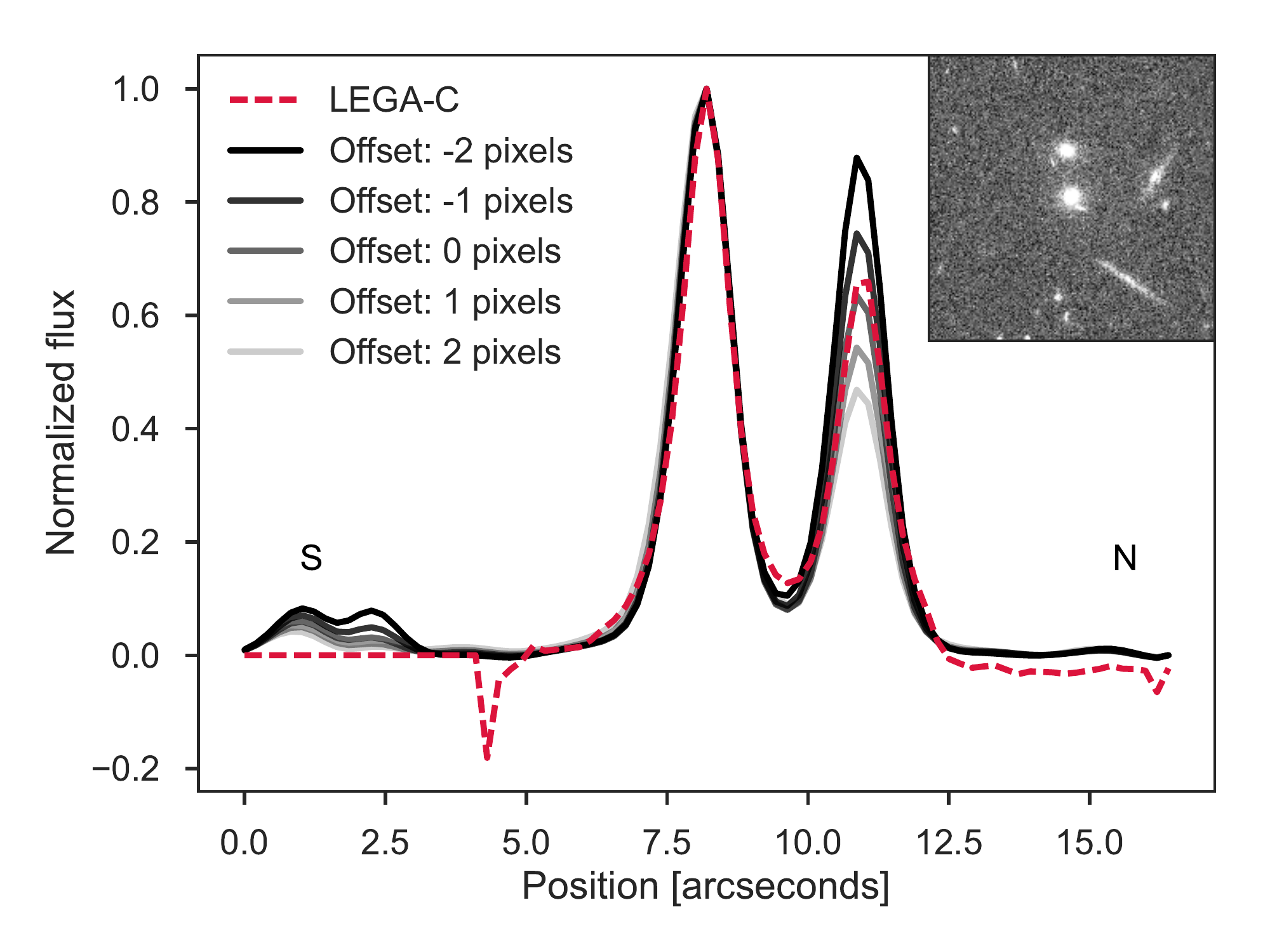}\label{fig:slit_placement}
\caption{Example of a galaxy with a close neighbour in the slit, visible and fittable in the light profile. In red-dashed, the normalized light profile obtained from collapsing the spectra in the wavelength direction. In successive greytones the lightprofile obtained by convolving the $HST/ACS$ image with $0.9$ Moffat seeing profile, rebinning, and placing a $5$ pixel slit (with pixels of $0.205$ arcseconds) with varying offsets. The inset shows the $HST/ACS$ image (North up, East left). The x-axis shows the distance from the bottom (South side) of the slit, in arcsec.}
\end{figure}

\renewcommand{\thefigure}{\arabic{figure}}

\end{document}